\documentclass[12pt]{article}
\usepackage{amsmath,amsthm,amssymb,euscript,epsf,epsfig,color,bbm}
\usepackage{array}
\usepackage{fancybox}
\usepackage{hyperref} 
\usepackage{comment}

\newcommand{\be}{\begin{equation}}
\newcommand{\ee}{\end{equation}}
\newcommand{\bea}{\begin{eqnarray}\displaystyle}
\newcommand{\eea}{\end{eqnarray}}

\makeatletter
\@addtoreset{equation}{section}
\makeatother

\def\one{{\hbox{ 1\kern-.8mm l}}}
\def\zero{{\hbox{ 0\kern-1.5mm 0}}}

\def\crit{ {\rm crit }}

\def\L{ L } 
\def\G{ G }

  \def\cF{{\cal F}}
\def\cG{{\cal G}}  \def\cI{{\cal I}}
  
 \def\cN{{\cal N}} \def\cO{{\cal O}}
  
 \def\cT{{\cal T}} 
 \def\cW{{\cal W}} 
 \def\cZ{{\cal Z}}

\def\PF{ { \rm PF } } 
 
\def\Sym{ {\rm{ Sym }} }

\def\Det{ {\rm\bf det} }
\def\Exponent{  { \rm Exponent } } 
\newtheorem{theorem}{Theorem} 

\def\mR{\mathbb{R}}

\begin{document}

\begin{flushright}
QMUL-PH-25-13 \\
\end{flushright}

\begin{center} 
{\large \bf   Gauged permutation invariant tensor quantum mechanics,  \\ 
 least common multiples and the inclusion-exclusion principle.
\\
 }

 \medskip

\bigskip

{\bf Denjoe O'Connor}$^{a,*}$, {\bf Sanjaye Ramgoolam}$^{b,\dag}  $

\bigskip

$^a${\em School of Theoretical Physics } \\
{\em Dublin Institute for Advanced Studies},\\
{\em 10 Burlington Road, Dublin 4, Ireland } \\
\medskip
$^{b}${\em  Centre for Theoretical Physics }\\
{\em Department of Physics and Astronomy}, \\
{\em Queen Mary University of London,} \\
{\em  London E1 4NS, United Kingdom }\\
\medskip
E-mails:  $^{*}$denjoe@stp.dias.ie,
\quad $^{\dag}$s.ramgoolam@qmul.ac.uk

\end{center} 

\begin{abstract} 
We derive the  canonical ensemble partition functions for gauged permutation invariant tensor quantum harmonic oscillator  thermodynamics, finding surprisingly simple expressions with number-theoretic characteristics. These systems have a gauged  symmetry 
of $S_N$, the symmetric group of all permutations of a set of $N$ objects. The symmetric group
acts on tensor variables $ \Phi_{ i_1, \cdots , i_s } $, where the $s$ indices  each range over $ \{ 1, 2, \cdots , N  \} $ and have the standard $S_N$ action of permutations. The result is a sum over partitions of $N$ and the summand is a product admitting simple expressions, which  depend on the least common multiples (LCMs)  of subsets of the parts of the partition.  The inclusion-exclusion principle of combinatorics plays a central role in the derivation of these expressions. 
The behaviour of these partition functions under inversion of the Boltzmann factor $ x = e^{ - \beta } $ is governed by universal  sequences associated with invariants of symmetric groups and alternating groups. The partition functions allow the development of a high temperature expansion analogous to the $s=2$ matrix case.  The breakdown point of the expansion leads to a critical Boltzmann factor  $ x_c = { \log N \over sN^{ s-1}}$  as the leading large N approximation for a transition in the model.

\end{abstract} 

\newpage 

\tableofcontents

 \section{ Introduction } 

Closed form expressions with number-theoretic characteristics for canonical ensemble partition functions for matrix models with gauged permutation  symmetry were derived in \cite{GPIMQM-PF}. 
The matrices $ \Phi_{ ij} $ transform as $V_N \otimes V_N$, where $V_{ N } $ is the standard representation of the symmetric group $S_N$. 
A powerful tool employed was the Molien-Weyl formula for the counting of invariants, which was derived from the path integral formulation of the gauged  permutation invariant matrix model in \cite{GPIMQM-PI}.  The Molien-Weyl formula gives the counting in terms of a sum over conjugacy classes of permutations of the inverse determinant of a permutation matrix, in the specified conjugacy class,  acting in the $V_N \otimes V_N$ representation. By using the explicit form of the eigenspectrum of the permutation matrix, which is easily calculable for this representation, the Molien-Weyl formula was simplified to give more tractable functions of the Boltzmann factor,  $x = e^{ - \beta \omega }$, where $ \beta = { 1 \over T } $ and $ \omega = \sqrt { k \over m } $ is the oscillator frequency. A detailed study of the thermodynamics of these models was undertaken in \cite{PIMQM-Thermo}. 
 These results built on the study of permutation invariant matrix models and matrix quantum mechanics initiated in \cite{LMT}\cite{PIGMM} \cite{PIMFac} \cite{PIMQM}. 

\vskip.2cm

The partition functions derived in  \cite{GPIMQM-PF} are sums over partitions of $N$, denoted by $p$,  of quantities 
$ \cZ  ( N , p  ,  x ) $, for which we derived a product formula (equation (2.14) in \cite{GPIMQM-PF}).  A high temperature expansion around $ x \sim 1 $ was developed in \cite{PIMQM-Thermo}. 
  By proving that the leading terms in this limit come from  partitions $p=[1^N] $ and $ p = [ 2,1^{N-2} ] $, a  value of the Boltzmann factor $ x_c \sim { \log N \over 2 N } $ was derived as the leading large $N$ approximation for the breakdown  of the high temperature expansion.  
The energy-temperature curve in the canonical ensemble shows a sharp transition at $ x = e^{ - \beta  \omega } \sim x_c = { \log N \over 2 N } $, while the transition region in the micro-canonical ensemble displays negative specific heat capacity below a critical energy $E_{ \crit } \sim {N \log N \over 2 }  $  \cite{PIMQM-Thermo}. These features of the thermodynamics were related to the very rapid growth of the degeneracy of states as a function of energy in the stable regime of state counting, where the number of quanta is less than $N$, along with the taming of this rapid growth by finite $N$ effects. When we keep $ \omega $ fixed, the $ x_c \rightarrow 0 $ behaviour in the large $N$ limit has the interpretation of a vanishing Hagedorn temperature. This was first identified in  the context of tensor models with continuous symmetries in \cite{Igor,Tseyt}. 

\vskip.2cm

The primary focus  of this paper is to generalise the mathematical expressions for canonical ensemble  partition functions from \cite{GPIMQM-PF} to the case of multi-index tensors 
$ \Phi_{ i_1 , i_2 , \cdots , i_s } $. Permutation invariant tensor models have been studied in \cite{PITM}. The counting of permutation invariant observables of degree $k$ was expressed in terms of partition algebras $P_{k}(N)$ and was also related to a graph counting problem, generalising the connections between graph counting and permutation invariant matrix observables found in \cite{PIGMM,PIG2MM}.  The graph-theory connection has also been useful in the application of permutation invariant matrix models for matrix data analysis \cite{GTMDS} \cite{PIMSCLT} \cite{PIGMF}. 

\vskip.2cm 

 In this paper we extend the calculation of canonical ensemble partition functions  in  \cite{GPIMQM-PF} to $s$-index tensors with gauged permutation symmetry. We uncover a beautiful number theoretic structure in these canonical partition functions, involving equations relating the LCMs of multiple integers to their GCDs. These relations between LCMs and GCDs are known to follow from the inclusion-exclusion principle of combinatorics \cite{WikiInEx}\cite{Rota}, and are used in arriving at our first main result \eqref{MainProps1}.  The inclusion-exclusion principle also plays another role in this paper, in going  from \eqref{MainProps1} to  a second formula \eqref{MainProps2}  for the $s$-dependent partition functions where a simple analytic structure of the $s$-dependence is  manifest. 

\vskip.2cm

The paper is organised as follows.  Section \ref{sec:mainresult} states our main results.  
In section \ref{sec:LCMMWdet} we review some relevant background material. We recall two key formulae from \cite{PIMQM-Thermo} which relate the calculation of Molien-Weyl determinants arising for $S_N$ to LCMs of integers. We also recall relevant mathematical background on the relation between the LCM of multiple numbers to products of GCDs (greatest common denominators)  of subsets of these numbers. 
In section \ref{sec:MWlcmgcd} we show that the Molien-Weyl determinants can be expressed as a product involving LCMs of parts of the partition $p$ of $N$ along with GCDs of the parts (equations \eqref{sindGL1}). In section \ref{LCMGCDtoLCM} we show that the LCM-GCD formula in \eqref{sindGL1}  can be further simplified to obtain the expression \eqref{MainProps1} in Theorem 1, which is purely in terms of LCMs.  In section \ref{sec:subsum} we show that the LCM formula in \eqref{MainProps1} can be simplified further to a form \eqref{subsetsumformula}  where the simple analytic structure as  a function of $s$ is manifest. In section \ref{limsdlts} we study the $ x \rightarrow \infty $ and the $ x \rightarrow 1 $ limits of the partition functions $ \cZ_s ( N , x )$. The behaviour for  $ x \rightarrow \infty $ benefits from an analysis of the transformation of the partition function under the transformation $ x \rightarrow x^{-1}$ and the outcome is that the large $x$ behaviour is sensitive to the first difference in the small $x$ expansion of the partition function for $S_N$ compared to that for the alternating group $A_N$.
Based on a well-motivated conjecture for the first two leading terms of the high temperature  expansion of $ \cZ_s ( N , x )$ as $ x \rightarrow 1$, we derive the leading large $N$ formula for the breakdown point $ x_c \sim  { \log N \over sN^{ s-1} } $ of the high temperature expansion. In section \ref{sec:thermodynamics} we discuss the thermodynamics of the permutation invariant tensor quantum models. This includes a characterisation of families of near-factorial degeneracies leading to negative specific heat capacities in the micro-canonical ensemble. 
 
\section{ Number-theoretic formulae for permutation invariant tensor model partition functions  } 
\label{sec:mainresult}

Consider  real variables $ \Phi_{ i_1 , i_2 , \cdots , i_s } $ where 
\bea 
 1  \le && i_1 ~~~  \le N \cr 
 1 \le  && i_2  ~~~ \le N \cr 
 && \vdots \cr 
 1 \le  && i_s ~~~ \le N  
\eea
They transform in the representation $V_N^{ \otimes s } $ where $V_N$ is the $N$-dimensional natural representation of $S_N$. Let $ \cZ_s ( N , k)$ be the dimension of the space of $S_N$ invariant polynomials of degree $k$ in the variables $ \Phi_{ i_1 , i_2 , \cdots , i_s } $. This is the dimension of the $S_N$ invariant subspace of the $k$-fold symmetric tensor power 
\bea 
\Sym^k ( V_N^{ \otimes s } ) 
\eea
which we denote 
\bea 
{\cal H}_{ S_N} (  \Sym^k ( V_N^{ \otimes s } )  )
\eea
with
\bea 
\cZ_s ( N , k)={\rm dim}({\cal H}_{ S_N} (  \Sym^k ( V_N^{ \otimes s } )  ))
\eea
The partition function $\cZ_s ( N , x )$ is the generating function of these dimensions 
\bea 
\cZ_s ( N , x ) = \sum_{ k =0}^{ \infty } x^{ k } \cZ_s ( N , k)
\eea
and the Molien-Weyl formula (a textbook reference for this formula with applications
 is \cite{MW}) gives 
\bea\label{sum1ovsymp} 
\cZ_s ( N , x ) && = { 1 \over N! } \sum_{ \sigma \in S_N }  { 1 \over \Det (  1 - x D^{ V_{ N }^{ \otimes s } } ( \sigma )   ) }  \cr 
&& =  \sum_{ p \vdash N } { 1 \over \Sym ~ p }  { 1 \over \Det (  1 - x D^{ V_{ N }^{ \otimes s } } ( \sigma_p  )   ) } 
\eea
We denote 
\bea\label{WeylFormp}  
\cZ_s ( N , p , x  ) 
= { 1 \over \Det (  1 - x D^{ V_{ N }^{ \otimes s } } ( \sigma_p  )   ) } 
\eea
For positive integers $a, b$, we will use the compact notation 
\bea\label{compLG}  
&& L  ( a , b ) = \hbox{LCM} ( a,b ) = \hbox { Least Common Multiple of $ a $ and  $ b $ } \cr   
&& G ( a , b ) = \hbox{GCD} ( a , b ) = \hbox { Greatest Common Divisor of $a$ and $b$ } 
\eea 
and similarly $ L (a_1 , a_2 , \cdots , a_n ) $ and $ G ( a_1 , \cdots , a_n)$ for the LCM and GCD of a sequence of positive integers. 

\subsection{ LCM formula for the Molien-Weyl inverse determinant } 
Let us present the partitions $p$ of $N$ as a list of non-zero numbers $a_i$, equipped with multiplicities $p_i >0 $, obeying $ \sum_{ i } a_i p_i = N $ so that $ p = \{ a_1^{ p_1} , a_2^{ p_2}  , \cdots , a_K^{ p_K} \} $ for some $ K \le N $.
We prove the following theorem \\
\begin{theorem}
The partition function $ \cZ_s  ( N , p ,  x)  $ which is the Molien-Weyl inverse determinant \eqref{WeylFormp} is equal to the following product depending on LCMs of numbers chosen from the set of cycle lengths $ \{ a_1 , a_2 , \cdots , a_K \}$ :
\bea\label{MainProps1} 
\cZ_s  ( N , p ,  x) = 
  \prod_{ i_1 \cdots i_s   \in \{ 1, \cdots , K \} } { 1 \over ( 1 - x^{ L ( a_{i_1} , \cdots , a_{ i_s} ) }  )^{ \frac{a_{i_1} \cdots  a_{ i_s}    }{L( a_{i_1},  
  \cdots ,  a_{i_s}  )  }    p_{i_1} \cdots p_{i_s}    }}
\eea
\end{theorem}
We will refer to the expression on the RHS as the LCM formula for $s$-index permutation invariant tensors. 

A slight variation is to present the partition $p$  as $ [ 1^{ p_1} , 2^{ p_2} , \cdots, N^{ p_N} ] $ where the $ p_i \ge 0$, and again $ \sum_{ i =1}^{ N } i p_i  = N$. 
\bea\label{MainProps2}
\cZ_s  ( N , p ,  x) =   \prod_{ i_1 \cdots i_s   \in \{ 1, \cdots , N \} } { 1 \over ( 1 - x^{ L ( i_1  , \cdots , i_s  ) }  )^{ \frac{  i_1\cdots   i_s    }{L(  i_1,  
  \cdots ,  i_s  )  }    p_{i_1} \cdots p_{i_s}    }}
\eea 
with the contribution to the product being 1 when the exponent in the denominator is zero.

For the $3$-index case, the theorem specializes to 
\bea\label{MainProp3}  
 \cZ_3  ( N , p ,  x) = 
  \prod_{ i ,  j ,  k  \in \{ 1, \cdots , K \} } { 1 \over ( 1 - x^{ L ( a_i ,   a_j , a_k  )  } )^{  {  a_i a_j a_k     \over L ( a_i ,  a_j , a_k)  } p_i  p_j p_k   }}
\eea
And for the $4$-index case
\bea\label{MainProp4}   
\cZ_4  ( N , p ,  x) = 
  \prod_{ i ,  j ,  k , l   \in \{ 1, \cdots , K \} }
     \frac{ 1 }{ ( 1 - x^{ L ( a_i ,   a_j , a_k , a_l  )  }  )^{
\frac{a_i a_j a_k a_l  }{L( a_i ,  a_j , a_k , a_l )  } p_i  p_j p_k  p_l      
      }       
     }
\eea

\subsection{ General  subset-sums LCM-formula } 

Consider partitions of $N$  presented as $ p = [ a_1^{ p_1 } , a_2^{ p_2} , \cdots , a_K^{ p_K } ] $, where $ p_1, p_2 , \cdots , p_K \ge 1$ and $ \sum_{ i = 1}^K a_i p_i = N$.  The formula \eqref{MainProps1} can be rewritten in terms of a sum over non-empty subsets $ S$ of $ \{ 1, 2, \cdots , K \} $. Such a subset $S$, of cardinality $  |S| \equiv l   $, takes the form $ \{ a_{ m_1 } , a_{ m_2} , \cdots , a_{ m_l } \} $ with $ m_1 , \cdots , m_l $ being distinct elements 
in $  \{ 1, 2, \cdots , K \}$. For  such  a subset $S$
let us define 
\bea\label{exp} 
L ( S ) = \hbox { LCM } ( a_{ m_1} , a_{ m_2} , \cdots , a_{ m_l } ) 
\eea
Further consider non-empty subsets $ H $ of $ S$. Such a subset of cardinality 
$ |H| = h   $ takes the form $ \{ n_1 , \cdots , n_h \} $ where $ n_1 , \cdots , n_h$ are distinct elements of $ \{ m_1 , \cdots , m_l \} $.  Define the function 
\bea\label{exps}  
W (N , p , s ;      S , H  )  =  \left (  \sum_{ j =1 }^{ h }  a_{ n_j } p_{ n_j } \right )^s  =   \left (  \sum_{ n  \in H  }  a_{ n } p_{ n } \right )^s
\eea  
and 
\bea\label{expss}  
\cW ( N , p , s ;  S ) = \sum_{  \substack { H  \subset S \\ H  \ne \emptyset   }  }  (-1)^{ |S| - |H| } W (N, p, s ;      S , H  ) 
\eea

\begin{theorem}\label{subsetsumtheorem} 
The partition function $ \cZ_s ( N , p , x  ) $, for the contribution of a partition $  p $ of $N$, for $S_N$ invariant $s$-index tensors, which is given by  \eqref{WeylFormp} and \eqref{MainProps1} can be re-written as 
\bea\label{subsetsumformula}  
\cZ_s ( N , p , x  ) =
 \prod_{ \substack {  S \subset \{ 1, \cdots , K  \} \\ S \ne \emptyset } }   
 { 1 \over ( 1 - x^{ L ( S )  } )^{ \cW ( N , p , s ; S ) \over L ( S )   }  }
\eea
with $ \cW ( N , p , s ;  S )$ as given in \eqref{expss}. 
\end{theorem}

\section{ Background on Least common multiples and Molien-Weyl determinants for $S_N$.   } 
\label{sec:LCMMWdet} 

In this section we collect useful mathematical results we will use later. A  simple but key formula in section  \ref{subsec:rtsLG} shows how a product over roots of unity leads to an expression in terms of the LCM and the GCD of a pair of integers.  In section \ref{subsec:LGmany} we review the known relation between the LCM of a list of integers and GCDs of various subsets of the integers. This relation is a consequence of the inclusion-exclusion principle \cite{WikiInEx}\cite{Rota} of combinatorics.

\subsection{ Two useful lemmas relating Molien-Weyl determinants to LCMs} 
\label{subsec:rtsLG} 

For some positive integer $a$ we denote the $a$'th root of unity 
\bea 
\omega_a \equiv  e^{ 2 \pi i \over a } \, .
\eea
The first useful and evident Lemma, which was useful in the permutation invariant matrix case in 
\cite{GPIMQM-PF}, is \\
\noindent 
{\bf Lemma 1: } 
\bea\label{Lemma1} 
\prod_{t=0}^{ a-1} {  ( 1 - x \omega_a^t )   }   = ( 1 - x^a ) 
\eea
The following is  Lemma eq. (4.15) from \cite{GPIMQM-PF}, and the elementary proof is given there
\noindent 
{\bf Lemma 2: } 
\bea\label{Lemma2} 
\prod_{t_1 = 0 }^{ a_1 -1 }  ( 1 - x^{ a_2} \omega_{ a_1}^{a_2 t_1}  ) 
=  ( 1 - x^{ L ( a_1 , a_2 )} )^{ G ( a_1 , a_2 ) } 
\eea

\subsection{ Useful mathematical facts about LCMs and GCDs  } 
\label{subsec:LGmany}

In this section, we review properties of LCMs and GCDs, following standard textbooks, the introductory chapter of \cite{Apostol}. We have also found this stack-exchange discussion to be useful \cite{GCDLCM-stack} along with the online resource \cite{cut-the-knot}.

\noindent 
{\bf  Standard relation between LCM and GCD for a pair of numbers } \\
The following identities hold 
\bea\label{Lab} 
 \L ( b_1 , b_2 ) = { b_1  b_2 \over  \G ( b_1 , b_2 ) } 
\eea 

\noindent 
{\bf Proof:  }  Consider the unique prime factorisations of the numbers $b_1 , b_2$, which exist by the fundamental theorem of arithmetic (e.g. Theorem 1.10 of \cite{Apostol}).  Let $ \PF  (b) $ be the set of prime factors of an integer $b$, and for a prime $p$ in this set, let $m_p ( b ) $ be the largest power of $p$ which divides $b$. 
\bea 
&& b_1 = \prod_{ p \in \PF ( b_1 )  } p^{ m_p ( b_1)    } \cr  
&& b_2 = \prod_{ p \in \PF ( b_2 )  } p^{ m_p  ( b_2  ) }  
\eea
Let us also denote $ \PF ( b_1 , b_2 ) \equiv \PF ( b_1 ) \cup \PF ( b_2 ) $. The relations between the LCM and the maxima of the exponents in the prime decomposition and between the GCD and the minima of the exponents in the prime decomposition are well known. 
\bea 
&& \G ( b_1 , b_2 ) = \prod_{ p \in \PF ( b_1 , b_2 ) }  p^{ \min  ( m_p  ( b_1 ) ,  m_p  ( b_2 ) )   } \cr 
&&  \L ( b_1 , b_2 ) = \prod_{ p \in \PF ( b_1 , b_2 ) } 
 p^{ \max  ( m_p  ( b_1 ) ,  m_p  ( b_2 ) )   } 
\eea
Now use 
\bea 
&& \max ( a , b  ) =  a ~ \cI  (  a \ge b ) + b  ~ \cI  ( a < b ) \cr 
&& \min ( a , b ) = b  ~ \cI  ( a \ge b ) + a  ~ \cI  ( a < b ) 
\eea
where the indicator function $ \cI  ( x ) $ is defined by: 
\bea 
&& \cI ( x ) = 1   \hbox { if $x$  holds }\cr  
&& \cI ( x ) = 0 \hbox {  otherwise }  
\eea

It follows that 
\bea 
 \max ( a , b  ) + \min ( a , b ) &=& a  ( \cI  ( a < b ) + \cI   (  a \ge b ) ) 
+ b  ( \cI ( a < b ) + \cI (  a \ge b ) )   \cr 
& = & a +b 
\eea
Applying this result to the prime decompositions 
\bea 
\G ( b_1 ) \L  ( b_2 ) &&  = \prod_{ p \in \PF ( b_1 , b_2 ) } p^{  \min ( m_p ( b_1 ) , m_p (b_2) ) 
+ \max (  m_p ( b_1 ) , m_p ( b_2 )     )     }  \cr 
&& = \prod_{ p \in \PF ( b_1 , b_2 )   }   
p^{ m_p ( b_1 ) + m_p ( b_2 ) }  = b_1 b_2 
\eea 
\noindent 
{ \bf Relation between LCM and GCD for a triple of numbers } 
\bea\label{Labc} 
 \L ( b_1 , b_2 , b_3 ) = {b_1 b_2 b_3  \G ( b_1 , b_2 , b_3 ) \over \G ( b_1 , b_2 ) G ( b_1 , b_3 ) \G ( b_2 , b_3 )} 
\eea 
We again use the prime decompositions to write 
\bea 
\L ( b_1, b_2 , b_3 ) && = \prod_{ p \in \PF ( b_1 , b_2 , b_3 ) } 
p^{ \max ( m_p ( b_1 ) , m_p ( b_2  ) , m_p ( b_3 ) )  } \cr 
\G ( b_1 , b_2 , b_3 ) && = \prod_{ p \in \PF ( b_1 , b_2 , b_3 ) } 
p^{ \min ( m_p ( b_1 ) , m_p ( b_2  ) , m_p ( b_3 )  ) }
\eea
The maximum and minimum of a triple of numbers is known to be related by 
\bea 
\max ( m_1 , m_2 , m_3 ) &=& m_1 + m_2 + m_3 - \min ( m_1 , m_2 ) - \min ( m_1 , m_3 )\\&&\qquad\qquad - \min ( m_2 , m_3 ) + \min ( m_1 , m_2 , m_3 ) \cr 
&&
\eea
An easy proof is to consider the sets 
\bea 
&& S_1 = \{ 1, 2, \cdots , m_1 \} \cr 
&& S_2 = \{ 1, 2, , \cdots , m_2 \} \cr 
&& S_3  = \{ 1, 2,  \cdots , m_3 \} 
\eea 
Observe that 
\bea 
\max ( m_1 , m_2 , m_3 )  = | S_1 \cup S_2 \cup S_3 | 
\eea
where the RHS is the cardinality of the union of the three sets, and 
\bea 
&& \min ( m_1 ,m_2 ) = | S_1 \cap  S_2 | \cr 
&& \min ( m_1 , m_2 , m_3 ) = | S_1 \cap  S_2 \cap S_3 |
\eea
We then use the inclusion-exclusion principle for sets which states that 
\bea 
&& | S_1 \cup S_2 \cup S_3 | =  |S_1| + |S_2| + | S_3| - |  S_1 \cap S_2 |  - |S_1 \cap S_3 | - | S_2 \cap S_3 | + |S_1 \cap S_2 \cap S_3 |  \cr 
&& 
\eea
For the statement of the inclusion-exclusion principle and its proof using indicator functions see \cite{WikiInEx}\cite{Rota}. 

\vskip.2cm 

\noindent 
{\bf General formula relating LCM and GCD for multiple positive integers } \\ 
\bea\label{GenLCMGCD}  
\L ( b_1 , \cdots  , b_n ) = ( b_1 \cdots b_n) \prod_{ k=2}^n  ( \cG_k ( b_1 , \cdots , b_n ) )^{ (-1)^{ k+1} }
\eea
where 
\bea 
\cG_k ( b_1 , \cdots , b_n ) 
=  \prod_{ i_1 < i_2 < \cdots < i_k \in \{ 1, \cdots , n \} } ( \G ( b_{ i_1}, b_{ i_2}  , \cdots , b_{ i_k } ) )
\eea
In other words, $ \cG_k $ is the product over $k$-element subsets, of the GCD for the subset. 

\noindent 
{\bf Proof:  } We use the prime decomposition of $b_i$ to relate the LCM and GCD's to $ \max ( m_1 , \cdots , m_n  ) $ and $ \min ( m_1 , \cdots , m_n ) $. 
The general statement of the inclusion-exclusion principle is 
\cite{WikiInEx} gives the relation. 
For the union of sets 
 \bea 
 W = \bigcup_{ i =1}^{ n  }  W_i 
 \eea
 we have an identity 
\bea 
| W | = \sum_{ k=1}^n (-1)^{ k+1} \left (  \sum_{ 1 \le i_1 < i_2 \cdots < i_k \in \{ 1, \cdots ,  n \}  }| A_{ i_1} \bigcap A_{ i_2} \cdots \bigcap A_{ i_k} |  \right ) 
\eea
Choosing $  | W_i | = m_i $, and $ W_{ i } = \{ 1, 2, \cdots , m_i \}$ gives 
the identity 
\bea 
\max ( m_1 , \cdots , m_n ) = \sum_{ k =1 }^{ n } (-1)^{k+1} \sum_{ i_1 < i_2 \cdots < i_k  \in \{ 1, 2, \cdots , n \} } \min  ( b_{ i_1} , b_{ i_2} , \cdots , b_{ i_k } ) 
\eea

\section{ From Molien-Weyl to LCM-GCD formula} 
\label{sec:MWlcmgcd} 

In this section, we prove that the Molien-Weyl formula for the
thermodynamic partition function $ \cZ_s ( N , p , x ) $ can be
expressed in terms of the LCMs and GCDs, of parts of the integer
partition $p$ of $N$. We refer to the result in equations
\eqref{sindGL1} as the LCM-GCD formula. The key technical tools here
are \eqref{Lemma1} and \eqref{Lemma2}, along with a description of the
eigenspaces of permutation operators in the natural (also called
standard) representation $V_N$ and its tensor powers $V_N^{ \otimes s
}$.

\subsection{  LCM-GCD formula for 3-index tensors   } 

Consider an integer partition $ p = [ a_1^{ p_1} , a_2^{ p_2} , \cdots , a_K^{ p_K} ] $ of $ N$,  with $a_i $ distinct positive integers in $ \{ 1, 2, \cdots , N \}$, $p_i$ positive integers $ 1\le  p_i \le N $ and $ \sum_i a_i p_i  = N$. With $L$ denoting the least common multiple and $G$ the greatest common divisor, we will prove the following result: 
\bea\label{MainProp}  
&& \cZ_s  ( N , p ,  x) = 
  \prod_{ i ,  j ,  k  \in \{ 1, \cdots , K \} } { 1 \over ( 1 - x^{ L ( a_i ,   a_j , a_k  )  } )^{  G ( a_i ,  a_j )  G ( L ( a_i , a_j ) , a_k )   p_i  p_j p_k  }   }\, .
\eea

\vskip.1cm 

\noindent 
{\bf Proof } \\
For a permutation $ \sigma $ with cycle structure given by $p$ as described above, the eigenvectors in the natural representation $V_N$  are 
\begin{align} 
&\omega_{ a_1}^{ t_1} ~~,~~  0 \le t_1 \le ( a_1 -1)   ~~; ~~  | \omega_{a_1}^{ t_1 } , r_1 \rangle ~~ 1 \le r_1 \le  p_1   \cr 
& \omega_{ a_2}^{ t_2} ~~,~~  0 \le t_2 \le ( a_2 -1)   ~~; ~~  | \omega_{a_2}^{ t_2 } , r_2 \rangle ~~ 1 \le r_2 \le  p_2   \cr 
& ~~ \vdots \cr 
& \omega_{ a_K}^{ t_K} ~~,~~  0 \le t_K \le ( a_K -1)   ~~; ~~  | \omega_{a_K}^{ t_1 } , r_K \rangle ~~ 1 \le r_K \le  p_K 
\end{align} 

In the 3-fold tensor product $V_N^{ \otimes 3 } $ the eigenvalues take the form 
\bea\label{mweig1}  
| \omega_{ a_{i_1} }^{t_{i_1} } , r_{1}  \rangle \otimes | \omega_{ a_{i_2} }^{t_{i_2}  } , r_{2}  \rangle \otimes 
| \omega_{ a_{i_3}}^{t_{i_3}  } , r_{3}  \rangle
\eea
The eigenvalues are 
\bea\label{mweig2}  
\omega_{ a_{i_1} }^{ t_{i_1} } \omega_{ a_{i_2} }^{ t_{i_2} } \omega_{ a_{i_3} }^{ t_{i_3} }  
\eea
with 
\bea\label{mweig3} 
&& 1 \le i_1 , i_2 , i_3 \le K \cr 
&& 0 \le t_{ i_1} \le ( a_{ i_1} -1 ) \cr 
&& 0 \le t_{ i_2} \le ( a_{ i_2} -1 ) \cr 
&& 0 \le t_{ i_3} \le ( a_{ i_3} -1 ) 
\eea
and the multiplicity space spanned by 
\bea\label{mweig4}  
&& 1 \le r_1 \le p_{ i_1}  \cr 
&& 1 \le r_2 \le p_{ i_2} \cr 
&& 1 \le r_3 \le p_{ i_3} 
\eea
has dimension $ p_{ i_1} p_{ i_2} p_{ i_3} $. 
Thus the determinant appearing in the Molien-Weyl formula is 
\bea 
&& { 1 \over \Det ( 1 - x D^{ V_N^{ \otimes 3 } } ( \sigma ) ) }  
 = \prod_{ i_1 =1}^{ K } ~ \prod_{ i_2 =1 }^K ~ \prod_{ i_3 =1}^{ K }   
\left (   \prod_{ t_{i_1}  = 0 }^{ a_{i_1}  -1 } ~ \prod_{ t_{i_2}  = 0}^{ a_{i_2}  -1 }  ~  \prod_{ t_{i_3}  = 0}^{ a_{i_3}  -1 } { 1 \over   ( 1 -  x\omega_{ a_{i_1} }^{ t_{i_1} } \omega_{ a_{i_2} }^{ t_{i_2} } \omega_{ a_{i_3}  }^{ t_{i_3} }   )^{ p_{ i_1} p_{ i_2} p_{ i_3}  }}  \right ) \cr 
&& 
\eea
It is useful to introduce the definition 
\bea 
\cF_{ 3 } ( b_1 , b_2 , b_3 ) =  \prod_{ t_1 = 0 }^{ b_1   -1 } ~ \prod_{ t_2 = 0}^{ b_2   -1 }  ~  \prod_{ t_3 = 0}^{ b_3   -1 } { 1 \over   ( 1 -  x\omega_{ b_1 }^{ t_1} \omega_{ b_2  }^{ t_2 } \omega_{ b_3  }^{ t_3}   )}
\eea
for the product over roots of unity. The inverse determinant is then 
\bea\label{ZF3form}  
{ 1 \over \Det ( 1 - x D^{ V_N^{ \otimes 3 } } ( \sigma ) ) }  
= \prod_{ i_1 =1}^{ K } ~ \prod_{ i_2 =1 }^K ~ \prod_{ i_3 =1}^{ K }   
   (  \cF_{ 3 } ( a_{ i_1}  , a_{ i_2}  , a_{ i_3 }  )  )^{ p_{ i_1} p_{ i_2} p_{ i_3} } 
\eea 

The next step  in the proof is to express the products over roots of unity in terms of LCMs and GCDs of the cycle lengths $a_i$ appearing in the cycle decomposition of the permutation 
\begin{align}
  \cF_{ 3 } ( b_1   , b_2   , b_3  )  ~ & =~\prod_{ t_1 = 0 }^{ b_1  -1 } ~ \prod_{ t_2 = 0}^{ b_2  -1 }  ~  \prod_{ t_3 = 0}^{ b_3  -1 } 
{ 1 \over ( 1 -  x\omega_{ b_1 }^{ t_1} \omega_{ b_2 }^{ t_2 } \omega_{ b_3  }^{ t_3}   ) } \cr 
~ &= ~ \prod_{ t_3 } \prod_{ t_2 }  { 1 \over ( 1 - x^{  b_1  } \omega_{ b_2 }^{ b_1  t_2} \omega_{ b_3 }^{ b_1 t_3} )  } \cr 
~ & = ~ \prod_{ t_3}  { 1 \over (   1 - x^{ L ( b_1  ,   b_2  )  }   \omega_{ b_3 }^{  L ( b_1  , b_2   ) .  t_3   } )^{  G ( b_1 , b_2  )  } }
\end{align} 
The second equality uses \eqref{Lemma1} to do the sum over $ t_1$. The third equality uses  \eqref{Lemma2}  to do the $t_2$ pproduct. 
In the next step we do  the $t_3$ product using \eqref{Lemma2} to get 
\bea\label{F3form} 
  \cF_{ 3 } ( b_1   , b_2   , b_3  )  && = { 1 \over 
(   1 - x^{ L ( L ( b_1  , b_2  )  , b_3 ) } )^{ G ( b_1  , b_2  )  G ( L ( b_1  , b_2 ) , b_3 ) }  } \cr 
&& = { 1 \over ( 1 - x^{ L ( b_1  , b_2  , b_3  ) } )^{ G ( b_1 , b_2   ) G ( L ( b_1  , b_2  ) , b_3 ) }  } 
\cr 
&& 
\eea
Combining \eqref{ZF3form} and \eqref{F3form} gives the proof of \eqref{MainProp}. \hfill $\blacksquare $. 

It is instructive to repeat the above derivation for the $4$-index case, where we start with 
\bea 
\cF_{ 4 } ( b_1 , b_2 , b_3  , b_4 ) =  \prod_{ t_1 = 0 }^{ b_1   -1 } ~ \prod_{ t_2 = 0}^{ b_2   -1 }  ~  \prod_{ t_3 = 0}^{ b_3   -1 }\prod_{ t_4 = 0}^{ b_4   -1 } { 1 \over   ( 1 -  x\omega_{ b_1 }^{ t_1} \omega_{ b_2  }^{ t_2 } \omega_{ b_3  }^{ t_3}  \omega_{ b_4}^{ t_4}   )}
\eea 
and arrive at 
\bea 
\cF_{ 4 } ( b_1 , b_2 , b_3  , b_4 ) = { 1 \over ( 1 - x^{ L ( b_1  , b_2  , b_3 , b_4   ) } )^{ G ( b_1 , b_2   ) G ( L ( b_1  , b_2  ) , b_3 )  G ( L ( b_1 , b_2 , b_3 ) , b_4 ) }  } 
\eea

\subsection{ The LCM-GCD formula for the $s$-index case } 

In this case, the calculation of the Molien-Weyl determinant, by using the spectrum of eigenvalues for $ \sigma \in S_N$ in the conjugacy class $ p = [ a_1^{ p_1} , \cdots , a_s^{ p_s} ] $ with $ p_1 , \cdots , p_s \ge 1 $ gives, using a straightforward generalisation of the steps  \eqref{mweig1}-\eqref{mweig4} of the $s=3$ case,  
\bea\label{MainProp}  
&& \cZ_s  ( N , p ,  x) = 
  \prod_{ i_1 , \cdots , i_s   \in \{ 1, \cdots , K \} }  ( \cF_{ s } ( b_1 , \cdots  , b_s ) )^{ p_{ i_1} \cdots p_{ i_s} }   \, .
\eea
where  
\bea\label{prodfrmsind} 
\cF_{ s } ( b_1 , \cdots  , b_s )  = \prod_{ t_1 = 0 }^{ b_1   -1 } \cdots \prod_{ t_s = 0}^{ b_s -1 } { 1 \over   ( 1 -  x\omega_{ b_1 }^{ t_1} \cdots  \omega_{ b_s}^{ t_s}   )}
\eea 

Application of the identity  \eqref{Lemma1} followed by repeated application
of \eqref{Lemma2} leads to 
\bea\label{sindGL1}  
  \cZ_s  ( N , p ,  x)= 
  \prod_{i_1 , \cdots , i_s \in \{ 1, 2, \cdots , K \}  }  { 1 \over ( 1 - x^{ L ( a_{ i_1} , a_{ i_2} , \cdots , a_{ i_s} } )  )^{( p_{ i_1} \cdots p_{i_s} ) \Exponent ( a_{ i_1} , \cdots , a_{ i_s} )   }}
  \eea
  \noindent
  where
\bea
\Exponent ( a_{ i_1}, \cdots , a_{ i_s}  )= G ( a_{ i_1}   ,  a_{ i_2} ) G ( L ( a_{ i_1}    ,  a_{ i_2}  ) , a_{ i_3} ) G ( L ( a_{ i_1}  , a_{ i_2}  , a_{ i_3} ) , a_{ i_4}  ) 
\cdots G ( L ( a_{ i_1}  , \cdots , a_{ i_{ s-1} }  ) , a_{ i_s }  )\nonumber 
\eea 
The proof is an evident iteration of the steps used in the case of the $3$-index  and $4$-index tensors.

\section{ From LCM-GCD formula to LCM-formula  }
\label{LCMGCDtoLCM} 

In this section we employ the background result detailed in section \ref{subsec:LGmany} to show that \eqref{sindGL1} is equivalent to \eqref{MainProps1}.

\subsection{ The general $s$-index tensor case }

For the $s$-index  case, the Molien-Weyl formula  leads \eqref{sindGL1}  to an exponent  of the form 
\bea\label{ExpfromMW}  
\Exponent ( b_1  , \cdots , b_s  ) = G ( b_1  , b_2 ) G ( L ( b_1   , b_2  ) , b_3) G ( L ( b_1 , b_2 , b_3 ) , b_4 ) 
\cdots G ( L ( b_1 , \cdots , b_{s-1} ) , b_{ s} ) \cr 
&& 
\eea
where the $b_j$'s are chosen from the set of parts of the partitions $ \{ a_1 , \cdots , a_K \}$. More compactly 
\bea\label{ExpfromMW1}  
\Exponent ( b_1  , \cdots , b_s  ) = \prod_{ l =1}^{ s-1} G ( L ( b_1 , \cdots , b_l ) , b_{ l+1} )
\eea 

{\bf Theorem 3 } We will prove that the exponent in \eqref{ExpfromMW} is equal to 
\bea\label{Thm3}  
\Exponent ( b_1  , \cdots , b_s  ) = { b_{ 1} b_{ 2} \cdots b_s \over L ( b_1 , b_2 , \cdots , b_s ) } 
\eea 

It will be useful to define 
\bea 
\cG_2 ( b_1 , \cdots , b_n ; b_{ n+1} ) =
 \sum_{ i  \in \{ 1 , \cdots , n \} } G ( b_i , b_{ n+1} ) 
\eea
\bea 
\cG_3 ( b_1 , \cdots , b_n ; b_{ n+1} ) = \sum_{ i < j \in \{ 1, \cdots , n \} } 
G ( b_i , b_j , b_{ n+1} ) 
\eea
and generally 
\bea 
\cG_k ( b_1 , \cdots , b_n ; b_{ n+1} ) 
= \sum_{ i_1 < i_2 \cdots < i_k \in \{ 1, \cdots , n \} } G ( b_{i_1} , b_{ i_2} , \cdots , b_{ i_k} , b_{ n+1} )  
\eea
In other words $\cG_k( b_1, b_2 , \cdots , b_n ; b_{ n+1} ) $  is the product over  $k$-element subsets  of $ \{ b_1 , \cdots , b_{ n+1} \}$, always including $b_{ n+1}$, of the GCD of the subset. 

We will first prove  an identity for the GCD of a pair consisting of 
the LCM of a sequence of numbers alongside a second number, i.e identities for 
$ G (  L ( b_1, b_2 , \cdots , b_n ) , b_{ n+1} ) $. The GCD of such pairs appear in \eqref{sindGL1}.  
The identity is \\
\noindent 
{\bf Proposition 1} \\ 
\bea\label{Prop1eq}  
&& G ( L ( b_1 , \cdots , b_n ) ,  b_{ n+1} ) 
= \prod_{ k =2}^{ n +1} \left (  \cG_{k} ( b_1 , \cdots , b_n ; b_{ n+1} ) \right )^{ \epsilon(k) }  \cr 
&& \epsilon(k) = (-1)^k  
\eea

\noindent 
{\bf Proof of Proposition 1 } \\ 
We start by using the relation \eqref{Lab}  between GCD, LCM and the product of a pair of numbers
\bea\label{fststpp1} 
&& G ( L ( b_1 , \cdots , b_n ) ,  b_{ n+1} )  = {  L ( b_1 , \cdots , b_n ) b_{ n+1} \over L ( L ( b_1 , \cdots , b_n ) , b_{ n+1} ) }  =  { L ( b_1 , \cdots , b_n ) b_{ n+1} \over L ( b_1 , b_2 , \cdots , b_n , b_{ n+1} ) } 
\eea
Now use \eqref{GenLCMGCD}
\bea 
L ( b_1 , \cdots , b_l ) = ( b_1 b_2 \cdots b_l ) \prod_{ k=2}^{ l}  ( \cG_{ k } ( b_1 , \cdots , b_l ) )^{  \epsilon (k +1 ) }  
\eea
This allows us to replace the LCMs in \eqref{fststpp1} with GCDs. 

This gives 
\bea 
&& G ( L ( b_1 , \cdots , b_n ) , b_{ n+1} )  
= ( b_1 \cdots b_{ n+1})
 \prod_{ k =2}^{n} ( \cG_{ k} ( b_1 , \cdots , b_n ) )^{ \epsilon ( k+1)  } \cr 
 && 
 \times { 1 \over ( b_1 \cdots b_{ n+1}) } \prod_{ k=2}^{ n +1 }  ( \cG_{ k} ( b_1 , \cdots , b_n , b_{ n+1}  ) )^{ \epsilon(k)  } 
\eea
In this expression, the GCDs of pairs, triples, in general size $k$-subsets, which do not contain $b_{ n+1} $ cancel out and we are left with an alternating product of GCDs which include $b_{ n+1}$. 
\bea\label{GCDdistLCM} 
G ( L ( b_1 , \cdots , b_n ) , b_{ n+1} )   
= \prod_{ k=2}^{ n+1} ( \cG_{ k} ( b_1 , \cdots , b_n  ; b_{ n+1} ) )^{ \epsilon(k)  } 
\eea
This completes the proof of the Proposition.  \hfill  $\blacksquare$

For the $s$-index  tensor partition function, we arrived at \eqref{ExpfromMW} which is not symmetric under permutations of the arguments. This initial symmetry of the product formula 
\eqref{prodfrmsind}  for the determinant was broken by choosing an order for doing the products. In the next proposition we will restore the symmetry. 

\noindent 
{\bf Proposition 2 } \\ 
\bea 
&& \Exponent ( b_1 , \cdots , b_s ) = \prod_{ k =2}^{s }  ( \cG_k ( b_1 , \cdots , b_s  ) )^{ \epsilon ( k ) } \cr 
&& \hbox{ where } \cG_k ( b_1 , \cdots , b_s ) = \prod_{ j_1 < j_2 \cdots < j_k \in \{ 1, \cdots , s \} } G ( b_1  , \cdots , b_k  ) \cr 
&& \hbox { and }  \epsilon ( k ) =  (-1)^{ k } ~~ \cr 
&& \hbox{ i.e. GCDs of even-size subsets appear in the numerator} \cr 
&& \hbox{ and GCDs of odd-size subsets appear in the denominator } 
\eea 

\noindent 
{\bf Proof  of Proposition 2 } \\ 
To understand the proof consider the following identity 
\bea 
\cG_2 ( b_1 , \cdots , b_s ) = \cG_2 ( b_1 , b_2 ) \cG_{ 2} ( b_1 , b_2 ; b_3 ) 
\cdots \cG_{ 2} ( b_1 , b_2 , \cdots , b_{s-1} ; b_{ s} ) 
\eea
The left hand-side is the product of GCDs of pairs of numbers chosen from $ \{ b_1 , \cdots , b_s \}$. On the RHS the last factor is the product of GCDs of pairs at least one of which is $b_s$, the second to last factor is the product of GCDs of pairs where at least one of the pair is $b_{ s-1}$ and the other is $b_i$ for some $ i < s-1$. In other words the pairs in $ \cG_2 (b_1 , \cdots , b_s ) $ are being ordered according to the highest number in the pair, and collected into products where the highest number is $b_2$, then the highest if $b_3$, etc. up to $b_{s}$. 
\bea 
\cG_{ 2} ( b_1 , \cdots , b_s ) 
= \prod_{ l=1}^{ s - 1} \cG_{ 2} ( b_1 , \cdots , b_l ; b_{ l+1} ) 
\eea
The same process of ordering and collecting for $k$-element subsets leads to 
\bea\label{cGandcH}  
\cG_{ k } ( b_1 , \cdots , b_s ) = 
\prod_{ l=1}^{ s -1  } \cG_{ k } ( b_1 , \cdots , b_l ; b_{ l+1} ) 
\eea

Now we can combine \eqref{ExpfromMW1}, \eqref{Prop1eq} to write the first two equalities  below 
\bea 
&& \Exponent ( b_1 , \cdots , b_s ) 
= \prod_{ l =1}^{ s-1} G ( L ( b_1 , \cdots , b_l ) , b_{ l+1} ) \cr 
&& = \prod_{ l=1}^{ s-1} \prod_{ k=2}^{ l+1} ( \cG_{ k} ( b_1 , \cdots , b_l  ; b_{ l+1} ) )^{ \epsilon(k) }  \cr 
&& = \prod_{ k=2}^{s}  \prod_{ l=1}^{ s-1} ( \cG_{ k} ( b_1 , \cdots , b_l  ; b_{ l+1} ) )^{ \epsilon(k)   } \cr 
&& = \prod_{ k=2}^s ( \cG_{ k} ( b_1 , \cdots , b_{ s } ) )^{ \epsilon( k )  } 
\eea
In the third line we re-arranged the product. In the fourth line, we used \eqref{cGandcH}. 

Finally we make one more use of \eqref{GenLCMGCD} to conclude that 
\bea
\Exponent ( b_1 , \cdots , b_s ) = 
{ b_{ 1} b_{ 2} \cdots b_s \over L ( b_1 , b_2 , \cdots , b_s ) } 
\eea
This completes the proof of the Theorem 3, equation  \eqref{Thm3}.  
Note that while the RHS of \ref{ExpfromMW} is not manifestly symmetric under permutations of 
$ \{ b_1 , \cdots , b_s \} $, the above result of  \eqref{Thm3} does have the symmetry.    \hfill $\blacksquare $

\section{ Subset-sum LCM formula for general  $s,N$ }
\label{sec:subsum} 

In this section we will derive our second main result \eqref{subsetsumformula} as given in Theorem \ref{subsetsumtheorem}.  We will give the general proof in section \ref{genproof}  by exploiting a useful result of wide applicability in combinatorics, called the  inclusion-exclusion principle. Curiously, the same principle underlies the relation between LCMs and GCDs of multiple integers, and this principle thus entered our derivation of the LCM-formula from the Molien-Weyl inverse determinant formula. It is perhaps a particularly striking example of the  wide applicability of the inclusion-exclusion principle that it has two distinct applications within the same paper. In the final section \ref{apps}, we give some special applications of the general result, which inform the discussion of the high temperature expansion in section \ref{limsdlts}. 

\subsection{ Derivation of the general subset-sums LCM-formula }\label{genproof} 

Consider the terms in \eqref{MainProps1} where $ i_1 , i_2 , \cdots , i_s $ all take values among some fixed subset $S$ of cardinality $ |S| = l$ which takes the form  $S = \{ m_1 , m_2 , \cdots , m_l \} $ with $ m_1, \cdots , m_l $ being distinct elements of  $ \{ 1, 2, \cdots , K \} $. 
These contribute a factor to the product. This factor contains $ LCM ( a_{ m_1} , \cdots , a_{ m_l } ) $ which we define as $ L (S ) $ for brevity. 
 Let $ t_{ 1} , t_2 , \cdots , t_l $ be  the number of the indices $ i_1 , \cdots , i_s $ which are equal to $ a_{ m_1} , a_{ m_2}, \cdots , a_{ m_l}   $ respectively. We will have a factor in the formula \eqref{MainProps1} which is 
\bea 
{ 1 \over ( 1 - x^{ L ( S ) }  )^{\cW  ( N , p , s ; S )   \over L ( S ) }}
\eea 
where 
\bea\label{sumforS}  
\cW  ( N , p , s ; S ) = 
\sum_{ \{ t_1 , t_2 , \cdots , t_l  \}   \in \Delta ( |S| ) } s!  \prod_{ j =1  }^l   
{ (a_{ m_j } p_{ m_j })^{ t_{ j } } \over   t_{ j } ! }  
\eea 
and $ \Delta ( |S | ) $ is the set of positive integer points in the simplex in $\mR^l$  specified by 
\begin{align} 
& t_{ \alpha } \ge 1  ~~ \hbox { for all } ~~ \alpha \in \{ 1, 2, \cdots , l \}  \cr 
& \sum_{ \alpha \in S } t_{ \alpha } = s 
\end{align} 
Taking the product over the different subsets  we have 
\bea\label{prodovS} 
\cZ_s ( N , p ,  x ) = \prod_{ S \subset \{ 1, \cdots , s \} } { 1 \over ( 1 - x^{ L ( S ) }  )^{\cW  ( N , p , s ; S )   \over L ( S ) }}
\eea
We will show that this exponent in \eqref{sumforS}, given as a sum over a simplex of dimension $ |S|$  is equal to the expression given in \eqref{expss}
as a signed sum over subsets of $S$.

The general proof uses the inclusion-exclusion principle \cite{WikiInEx}\cite{Rota}, expressed as an equality of indicator functions, or equivalently in more physical notation, as an equality of delta functions. Consider  the  set of  non-negative integer points $ \{ t_1 , t_2 , \cdots , t_l \}$ in  $ \mR^{ l }  $ obeying: 
\begin{align} 
& t_1 + t_2 + \cdots + t_l = s \cr 
&0 \le t_1 \le s \cr 
& 0 \le t_2 \le s \cr 
& ~~~~ \vdots \cr 
& 0 \le  t_l \le s  
\end{align}
We will call this the set $ \cT $. The summand in \eqref{sumforS} will be denoted 
$ f( t_1 , t_2 , \cdots , t_l )$ : 
\bea 
f( t_1 , t_2 , \cdots , t_l ) = s!  \prod_{ j =1  }^l   
{ (a_{ m_j } p_{ m_j })^{ t_{ j } } \over   t_{ j } ! } 
\eea  
 We start with the observation that 
\bea\label{simpsum}  
\sum_{ \vec t   \in \cT} f ( t_1 , t_2 , \cdots , t_l ) = ( a_{m_1}  p_{m_1} + a_{m_2} p_{m_2} + \cdots + a_{m_l}  p_{m_l}  )^s  = \left ( \sum_{ m \in S } a_{ m} p_{ m } \right )^{ s } 
\eea 
For each $ b \in \{ 1 , 2, \cdots , l \} $ define 
$ \cN_b $ to be the subset of $\mR^l $ with $ t_b =0 $.
The sum in \eqref{sumforS} is the sum of $ f ( t_1 , t_2 , \cdots , t_l ) $  over the complement in $ \cT $ of  $ \cT \cap \cN $ where 
\bea 
\cN =  \cN_1 \cup \cN_2 \cdots \cup \cN_l 
\eea
Note that $ \cN_1 \cap \cN_2 \cdots \cap \cN_l $ has zero intersection with $ \cT$ because of the first condition in the definition of $ \cT$, i.e. setting all the $ t_1 , \cdots , t_l $ to zero means that they cannot sum to $s$. However any subset of the $t$'s can be set to zero. 
Thus 
\bea\label{cTComp}  
W ( N , p , s ; S  ) = \sum_{ \vec t \in \cT } f ( t_1 , t_2 , \cdots , t_l )  
- \sum_{ \vec t \in  ( \cN \cap \cT)  } f ( t_1 , t_2 , \cdots , t_l )  
\eea 
The term being subtracted can be written as 
\bea\label{termSubt} 
\sum_{ \vec t \in \cN \cap \cT  } f ( t_1 , t_2 , \cdots , t_l )   = 
\sum_{ \vec t   \in \cT  } f ( t_1 , t_2 , \cdots , t_l ) \delta ( \vec t ,  \cN ) 
\eea 
Now use the fact that the delta function over the union  $ \cN $ of sets $ \cN_1 , \cN_2 , \cdots  , \cN_l  $ is equal, by the inclusion-exclusion principle \cite{WikiInEx,Rota}, to a signed  sum of delta functions over intersections 
of the subsets of $ \cN_1 , \cN_2 , \cdots , \cN_l$. These subsets are parameterised by subsets 
$J \subset \{ 1, 2, \cdots , l \}$ of cardinality $ |J| \equiv  j $ in the range $ 1 \le j \le l $. Such a subset takes the form $ \{ b_1 , \cdots , b_j \} $ and there is a corresponding intersection which we denote as $ \cN_J$ 
\bea 
\cN_{ J } \equiv \cN_{ b_1} \cap \cN_{ b_2} \cdots \cap \cN_{ b_j } 
\eea
The expansion of the delta function over $ \cN$ in terms of the delta function over the subsets given the inclusion-exclusion principle is 
\bea\label{InExdelts}  
\delta ( \vec t ,  \cN )  
=  \sum_{ j =1  }^{ l }  \sum_{ \substack { J \subset \{ 1 , \cdots , l \}  \\ |J| = j } } ( -1)^{ j -1 }  \delta ( \vec t , \cN_{ J  } )  
\eea
For  $ J   \subset \{ 1, 2, \cdots ,l \} $, 
\bea\label{InExdelts1}  
&& \sum_{ \vec t  \in \cT  }  f ( t_1 , t_2 , \cdots , t_l ) ~~ \delta ( \vec t , \cN_J ) 
=  \sum_{ \substack{ \{ t_{ c } \} \in \cT \\  c \in \{ 1 , \cdots , l \}  \setminus J }  }  s!  \prod_{ c \in \{ 1 , \cdots , l \}  \setminus J  } 
{   ( a_{ c  } p_{ c } )^{ t_{ c } }  \over  t_{ c }! }  \cr 
&& =  \left ( \sum_{ c \in \{ 1 , \cdots , l \} \setminus J }  a_{ c } p_{ c } \right )^s 
\eea
Define $ H = \{ 1 , 2, \cdots , l \}  \setminus J $ to be the complement of $J$ in $\{ 1, 2, \cdots , l \}$, and $ |H| = h $, so that $ h = ( l - j)  $ and 
\bea 
(-1)^{ j -1 } = (-1)^{ l - h  -1 } 
\eea
When $ j = l$ we have the $ \cN_J = \cN_1 \cap \cN_2 \cdots \cap \cN_l $, which as noted earlier in the proof, has zero intersection with $ \cT $.  Therefore, when  we use the expansion \eqref{InExdelts} in the LHS of \eqref{InExdelts1}, we can restrict the sum over $j$ to the range from $1$ to $(l-1)$. 
We thus get  a sum  over  all subsets $H \subset \{ 1, 2, \cdots , l \}  $ of cardinality ranging from $ 1 $ to $ l-1$. Thus \eqref{termSubt} can be written as 
\bea\label{newtermSubt}  
&& \sum_{ \vec t   \in \cT  } f ( t_1 , t_2 , \cdots , t_l ) \delta ( \vec t ,  \cN )   \cr 
&& = \sum_{ h = 1 }^{ l -1  } (-1)^{ l- h - 1 }  \sum_{ \substack {  H \subset \{ 1, \cdots , l \} \\ |H| = h  } } \left ( \sum_{ i \in H } a_{ i  } p_{ i } \right )^s 
\eea 
Using \eqref{cTComp} along with \eqref{simpsum} and  \eqref{newtermSubt}  we have 
\bea
\cW ( N , p , s ; S ) = \left ( \sum_{ \alpha \in S } a_{ \alpha } p_{ \alpha } \right )^{ s }  
+ \sum_{ h = 1 }^{ l -1  } (-1)^{ l- h }  \sum_{ \substack {  H \subset [l] \\ |H| = h  } } \left ( \sum_{ i \in H } a_i  p_{ i } \right )^s 
\eea  
We can absorb the first term into the second by extending the sum over $h$ up to $ l$ 
\bea\label{sumHsubS}  
\cW ( N , p , s; S   ) = \sum_{ h = 1 }^{ l   } (-1)^{ l- h }  \sum_{ \substack {  H \subset [l] \\ |H| = h  } } \left ( \sum_{ i  \in H } a_{ i } p_{ i  } \right )^s 
\eea 
Recalling \eqref{prodovS},  this completes the proof of {\bf Theorem}  {\bf\ref{subsetsumtheorem}}.  

\subsection{ Applications }\label{apps} 

Consider the case $ p = [ d^{ N/d} ] $ for any divisor of $N$. To apply Theorem \ref{subsetsumtheorem} we consider the set $ S = \{ 1 \} $ with $a_1 = d $.  The unique non-empty subset of $S$ is $S$ itself. The product over subsets in \eqref{subsetsumformula} collapses to a single term. The LCM of the subset is 
\bea 
L ( S ) = d 
\eea
The sum over subsets $H$ in \eqref{expss} also collapses to a single term. Evaluating the terms in \eqref{exps} and \eqref{exp} we thus have
Thus 
\bea 
\cZ_s ( N , p = [ d^{ N/d} ] , x ) = { 1 \over ( 1 - x^d )^{ N^s \over d } }
\eea 
In particular for $ d =1 , p = [ 1^N] $, we have 
\bea 
\cZ_s ( N , p = [ 1^{ N} ] , x ) = { 1 \over ( 1 - x )^{ N^s} } 
\eea
which we argue to be the leading term in the high temperature expansion in section \ref{limsdlts}. 

The next family of examples we consider is $  = [ 1^{ N - ar } , a^r ] $. 
In this case, the application of Theorem 2 involves working with the set  of cycle lengths 
$ \{ 1 , a \}$. The associated  set of indices specifying the distinct cycle lengths is
 $ \{ 1 , 2 \} $, with $ a_1 = 1 , a_2 = a $. Subsets $S$  are $ \{ \{ 1, 2 \} , \{ 1 \} , \{ 2 \} \} $, and the corresponding subset-LCMs $L(S)$ are $ \{ L ( 1 ,a )  = 1 , L ( 1 ) = 1 , L ( a ) = a \} $. From $ S = \{ 1 , a \} $ we have subsets $ H $ which can be $ \{ 1,2 \} , \{ 1 \} , \{ 2 \} $. Evaluating  the sum over subsets $H$ in \eqref{exps}  leads to the factor 
 \bea 
 { 1 \over ( 1 - x^a )^{ { 1 \over a  } (  N^s - ( N - ar)^s - (ar)^s   ) } }
 \eea 
 \eqref{expss}.  For $S = \{ 1 \} $ we have $ L ( S ) = 1 $ and $H=S$, so that \eqref{expss} evaluates to 
 \bea 
 { 1 \over ( 1 - x )^{ ( N - ar)^s  }  }
 \eea 
 For $  S = \{ 2 \} $, we have $ L(S) = a $, and $  H = S $ so that \eqref{expss} evaluates to 
 \bea 
 { 1 \over ( 1 - x )^{ { 1 \over a } (ar)^s }}
 \eea 
 Multiplying the factors 
\bea\label{1anda}  
\cZ_s ( N , p = [ 1^{ N-ar } , a^r ] , x ) 
= { 1 \over ( 1-x)^{ (N -ar)^s } } { 1 \over ( 1- x^a )^{ { 1\over a } ( N^s - ( N - ar )^s ) } } 
\eea
Note that the exponent of $ ( 1- x^a)$ is an integer. This is evident  from  the expression  \eqref{MainProps1} of Theorem 1 (which has been re-expressed in Theorem 2) that the exponents ${  a_{ i_1 } \cdots a_{ i_s}  \over L   ( a_{i_1} , \cdots , a_{i_s} ) } $ are  integers.

We will also argue that the first sub-leading term in the high-temperature expansion is a special case of \eqref{1anda} 
\bea 
\cZ_s ( N , p = [ 1^{ N-2 } , 2 ] , x)  = 
 { 1 \over ( 1-x)^{ (N -2)^s } } { 1 \over ( 1- x^2 )^{ { 1\over 2} ( N^s - ( N - 2 )^s ) } } 
\eea

Now, as a more intricate application of Theorem 2,  consider
the partition function $ \cZ ( N , p , s ; x ) $, for $ p = [  1^{ N - a_2 r_2 - a_3 r_3 } , a_2^{ r_2} , a_3^{ r_3 } ] $.
Here we have a set of cycle lengths $ \{ a_1 = 1, a_2 , a_3 \}$, cycle length multiplicities $ \{  p_1 = N - a_2 r_2 - a_3 r_3 , p_2 = r_2 , p_3 = r_3  \} $, and associated  indices 
$ \{ 1, 2, 3 \} $ identifying the distinct cycle lengths.   The subsets  $S$ of the  index set 
 are $  \{ \{ 1 \} , \{ 2 \} , \{ 3 \} , 
 \{ 1,2 \} , \{ 1, 3 \} , \{ 2, 3 \}, \{ 1,2,3 \} \} $.  For $ S = \{ 1 \} $, $ L ( S ) = 1 $. 
 The factor is 
 \bea 
 { 1 \over ( 1 - x )^{  ( N - a_2 r_2 - a_3 r_3 )^s   }  } 
 \eea 
 For $ S = \{ 2 \} $ 
 \bea 
 { 1 \over ( 1 - x^{ a_2} )^{  { 1 \over a_2}  ( a_2 r_2 )^s  } }
 \eea
 For $ S = \{ 3 \} $ 
 \bea 
 { 1 \over ( 1 - x^{ a_3 } )^{ { 1 \over a_3 } ( a_3 r_3 )^s } } 
 \eea
For $ S = \{ 1, 2 \} $, $ L ( S ) = a_2$. For $ H = S$, $ ( a_1 p_1 + a_2 p_2)^s = ( ( N - a_2 r_2 - a_3 r_3 ) + a_2 r_2   )^s   = ( N - a_3 r_3 )^s  $. For $ H = \{ 1 \} $,   $ (a_1 p_1)^s  
= ( N - a_2 r_2 - a_3 r_3)^s $. For $ H = \{ 2 \} $, we have $ ( a_2 p_2)^s  = ( a_2 r_2 )^s $. Collecting terms 
\bea 
 { 1 \over ( 1 - x^{ a_2 } )^{ { 1 \over a_2 }  ( (   N - a_3 r_3  )^s  - ( N - a_2 r_2 -  a_3 r_3 )^s   - ( a_2 r_2 )^s  )  }   } 
\eea 
Similarly for $ S  = \{ 1,3 \} $, 
\bea 
 { 1 \over ( 1 - x^{ a_3 } )^{ { 1 \over a_3 }  ( (   N - a_2 r_2  )^s  - ( N - a_2 r_2 -  a_3 r_3 )^s   - ( a_3 r_3 )^s  )  }   } 
\eea 
For $ S = \{ 2,3 \} $
\bea 
 { 1 \over ( 1 - x^{ L ( a_2 ,  a_3 )  } )^{ { 1 \over  L ( a_2 , a_3 )  }  ( (   a_2 r_2 + a_3 r_3  )^s  - ( a_2 r_2  )^s   - ( a_3 r_3 )^s  )  }   } 
\eea 
For $ S = \{ 1, 2, 3 \} $, 
\bea 
\cW ( N , p , s ;  x ) & = &   N^s - ( N - a_3 r_3)^s - ( N - a_2 r_2 )^s - ( a_2 r_2 + a_3r_3 )^s     \cr 
&&  +  ( N - a_2 r_2 - a_3 r_3)^s + ( a_2 r_2)^s + ( a_3 r_3 )^s   \cr 
&& 
\eea 
with each term being associated with a subset $H \subset S$ according to the formula \eqref{exps},  in the order of the terms presented above the subsets $H$  are $ \{ 1,2, 3 \} , \{ 1 , 2 \}  , \{ 1 , 3 \} , \{ 2 , 3 \} , \{ 1 \} , \{ 2 \} , \{ 3 \} $
\bea 
&& { 1 \over ( 1 - x^{ L ( a_2 ,  a_3 ) } )^{ { 1  \over L ( a_2  , a_3 ) }  (  N^s - ( N - a_3 r_3)^s - ( N - a_2 r_2 )^s - ( a_2 r_2 + a_3r_3 )^s      +  ( N - a_2 r_2 - a_3 r_3)^s + ( a_2 r_2)^s + ( a_3 r_3 )^s  )  }   }  \cr 
&&
\eea 
Collecting all the factors, one for each subset $S$ of the  cycle-length indexing set,  as per \eqref{subsetsumformula}
\bea 
&& \cZ_s ( N , p = [ 1^{ N - a_2 r_2 - a_3 r_3  } , a_2^{ r_2} , a_3^{ r_3 } ] , x ) \cr 
&& =  { 1 \over ( 1 - x )^{ ( N - a_2 r_2 - a_3 r_3 )^s   } }   { 1 \over   ( 1 - x^{ a_2 }  )^{ { 1  \over a_2 }  ( ( N - a_3 r_3   )^s - ( N - a_2 r_2 - a_3 r_3 )^s  )   }    }   { 1 \over   ( 1 - x^{ a_3 }  )^{ { 1  \over a_3 } (  ( N - a_2 r_2   )^s - ( N - a_2 r_2 - a_3 r_3 )^s  )   }    } \cr 
&& { 1 \over ( 1 - x^{ L ( a_2 ,  a_3 ) } )^{ { 1  \over L ( a_2  , a_3 ) }  (  N^s - ( N - a_3 r_3)^s - ( N - a_2 r_2 )^s     +  ( N - a_2 r_2 - a_3 r_3)^s  )  }   } 
\eea

\section{ Limits and dualities for $S_N$ invariant tensor partition functions }
\label{limsdlts} 

In this section, we study the limits $ x \rightarrow \infty $ and 
$ x \rightarrow 1$ of the partition funtion $ \cZ_s ( N , x )$.

\subsection{ Behaviour at  $x \rightarrow \infty $   }

Our discussion of the Molien-Weyl formula for the canonical partition function of the quantum mechanics of $s$-index tensors $ \Phi_{ i_1 , i_2 , \cdots , i_s } $  having the standard harmonic oscillator potential $  { 1 \over 2 } \sum_{ i_1 , \cdots , i_s =1}^N  ( \Phi_{ i_1 \cdots i_s } )^2 $ is easily generalised to the case of $d$ independent copies of these 
tensors. In this case the quantum mechanical variables are
  $ \Phi^{(a)}_{ i_1 , i_2 , \cdots , i_s } $ with   $ 1 \le  a \le d $, and the refined partition counts states with specified numbers of oscillators for each  species of  tensor. 
It will be instructive to consider the $S_N$ invariant partitions generalised to this case as we study the $ x \rightarrow \infty  $, and more generall $ x_a \rightarrow \infty $.  
\begin{equation}\label{d-G-s-tensor Result}
  Z_{G}(x_1,\cdots,x_d,s)=\frac{1}{\vert G\vert}\sum_{g\in G}\prod_{a=1}^d\frac{1}{{\rm\bf det}[1-x_a g^{\otimes s}]}\,.
  \end{equation}
where $g^{\otimes s}=g\otimes g\otimes\cdots\otimes g$, i.e. the $s$-fold tensor product of $g$.
The derivation of the Molien-Weyl formula from the gauged finite group quantum mechanical path integral was given in \cite{GPIMQM-PI}.  Our primary focus here continues to be $ G = S_N$, although the alternating group $A_N$ will also make an interesting appearance. 

To obtain the behaviour as $x_a\rightarrow\infty$ we can extract $-x_a g^{\otimes s}$ from the determinant in (\ref{d-G-s-tensor Result}) and changing the sum on $g$ to $g^{-1}$ and with ${\rm \bf det}[g^{-1}]={\rm \bf det}[g]^{-1}$ we obtain
\begin{equation}\label{d-G-s-tensor Result}
  Z_{G}(x_1,\cdots,x_d,s)=\frac{1}{\vert G\vert}\sum_{g\in G}\prod_{a=1}^d{{\rm \bf det}[-x_a^{-1} g^{\otimes s}]}\frac{1}{{\rm\bf det}[1-x^{-1}_a g^{\otimes s}]}\,.
\end{equation}

To further analyse this high temperature limit we need the result
\begin{equation}
  {\rm \bf det}[A\otimes B]={\rm \bf det}[A]^{\rm dim(B)}{\rm \bf det}[B]^{\rm dim(A)}\label{dettensorprod}
  \end{equation}
Factoring the product as
$$A\otimes B=(A\otimes \mathbbm{1}_{\rm dim B}).(\mathbbm{1}_{\rm dim A}\otimes B)$$
and noting that
$${\rm \bf det}[A\otimes B]={\rm \bf det}[A\otimes \mathbbm{1}_{\rm dim B}]{\rm \bf det}[\mathbbm{1}_{\rm dim A}\otimes B]$$
we have a block structure of repeated $A$ or $B$ so that 
$${\rm \bf det}[A\otimes \mathbbm{1}_{\rm dim B}]={\rm \bf det}[A]^{\rm dim_B}\qquad {\rm \bf det}[\mathbbm{1}_{\rm dim A}\otimes B]={\rm \bf det}[B]^{\rm dim_A}$$
establishing the result (\ref{dettensorprod}).

Returning to (\ref{d-G-s-tensor Result}) using (\ref{dettensorprod}) with $g$ of dimension $N$ we have
$${\rm \bf det}[g\otimes \mathbbm{1_N}^{\otimes{s-1}}]={\rm \bf det}[g]^{N^{s-1}}$$
which occurs for each of the $s$ factors in the decomposition of $g^{\otimes s}$. We therefore have
\begin{equation}
  {\rm \bf det}[\lambda g^{\otimes s}]=\lambda^{N^s}{\rm \bf det}[g]^{s N^{s-1}}
\end{equation}
with $\lambda$ a scalar factor multiplying the tensor product.
We therefore find
\begin{equation}
  \prod_{a=1}^d{{\rm \bf det}[-x_a^{-1} g^{\otimes s}]}=(-)^{dN^s}(\prod_{a=1}^dx_a^{-1})^{N^s}{{ { \rm \bf det}[g]^{s d N^{s-1}}}}
  \end{equation}

Then (\ref{d-G-s-tensor Result}) gives
\begin{equation}\label{d-G-s-tensor-x-inversion}
  Z_{G}(x_1,\cdots,x_d,s)=\frac{(-)^{dN^s}(\prod_{a=1}^dx_a)^{-N^s}}{\vert G\vert}\sum_{g\in G}{{\rm \bf det}[g]^{s d N^{s-1}}}\prod_{a=1}^d\frac{1}{{\rm\bf det}[1-x_a^{-1} g^{\otimes s}]}\,.
\end{equation}
So if ${{\rm \bf det}[g]^{s d N^{s-1}}}=c(N,d,s)$ independent of $g$ we have
the relation between the large $x_a$ and the small $x_a$ partition functions
\begin{equation}
\frac{Z_{G}(x_1,\cdots,x_d,s)}{Z_{G}(x_1^{-1},\cdots,x_d^{-1},s)}={(-)^{dN^s}(\prod_{a=1}^dx_a)^{-N^s}}c(N,d,s)
  \end{equation}
This in turn leads to a palindromic relation for the polynomial numerator of $Z_{G,s}(x,d,N)$.

The determinant in the Molien-Weyl formula only depends on the conjugacy class and specialising to symmetric group $G=S_{N}$
eq. (\ref{d-G-s-tensor Result}) becomes

\begin{equation}\label{d-SN-s-tensor Result}
  Z_{S_N}(x_1,\cdots,x_d,s)=\frac{1}{N!}\sum_{p\vdash N}\frac{1}{{\rm Sym} ~ p}\prod_{a=1}^d\frac{1}{{\rm\bf det}[1-x_a g_p^{\otimes s}]}\,.
  \end{equation}
where $\vert G\vert=N!$, $g_p$ is a representative of the p-th conjugacy class and
\begin{equation}
{\rm Sym} ~ p = \prod_{i=1}^n a_i^{p_i}p_i! 
\end{equation}
where we have used the form $p =[a_1^{p_1},a_2^{p_2},\cdots, a_{K}^{p_K}]$, with 
$a_i$ distinct non-zero parts (cycle lengths) such that $1\le a_i \le N $ and $p_i$ are positive integers ordered as $a_1 < a_2 < \cdots < a_K$ and 
$N=\sum_{i =1}^K a_i p_i$. 

The inversion formula (\ref{d-G-s-tensor-x-inversion}) now becomes
\begin{equation}\label{d-G-s-tensor-x-inversion}
Z_{S_N}(x_1,\cdots,x_d,s)=\frac{(-)^{dN^s}(\prod_{a=1}^dx_a)^{-N^s}}{N!}\sum_{p\vdash N}\frac{1}{{\rm Sym} ~ p}{{\rm \bf det}[g_p]^{s d N^{s-1}}}\prod_{a=1}^d\frac{1}{{\rm\bf det}[1-x_a^{-1}  g_p^{\otimes s}]}\,.
  \end{equation}
Since for any conjugacy class ${\rm \bf det}[g_p]=\pm 1$, i.e. the parity of
the partition we can split the partition into contributions from positive and negative parity defining
\begin{equation}\label{d-SN-s-tensor-with-parity}
  Z^{\pm}_{S_N}(x_1,\cdots,x_d,s)=\frac{1}{N!}\sum_{p\vdash N}\frac{1}{{\rm Sym}~p}\frac{1\pm {\rm \bf det}[g_p]}{2}\prod_{a=1}^d\frac{1}{{\rm\bf det}[1-x_a^{-1}  g_p^{\otimes s}]}\,.
  \end{equation}
Note that since the sign ${\rm \bf det}[g_p]$ has a fixed sign on a given parity sector so does ${\rm \bf det}[g_p]^{s d N^{s-1}}$, it is positive when $s d N^{s-1}$ is even and negative when odd. We therefore have the inversion result
\begin{equation}
Z^{\pm}_{S_N}(x_1,\cdots,x_d,s)=(-1)^{d N^s}(\pm1)^{s d N^{s-1}}(\prod_{a=1}^dx_a)^{-N^s}Z^{\pm}_{S_N}(x_1^{-1},\cdots,x_d^{-1},s)
\end{equation}
obtained  by splitting into positive parity
partitions and negative parity partitions.  
Therefore when $s d N^{s-1}$ is even both parities contribute positively to the inversion and the full partition function satisfies the inversion relation
\begin{equation}
Z_{S_N}(x_1,\cdots,x_d,s)=(-1)^{dN^s} ~( \prod_{a=1}^dx_a)^{-N^s}(Z_{S_N}(x_1^{-1},\cdots,x_d^{-1},s)
\end{equation}
When  $s d N^{s-1}$ is odd we have
\begin{equation}
Z_{S_N}(x_1,\cdots,x_d,s)=(-1)^{dN^s}(\prod_{a=1}^dx_a)^{-N^s} ( Z^{+}_{S_N}(x_1^{-1},\cdots,x_d^{-1},s)-Z^{-}_{S_N}(x_1^{-1},\cdots,x_d^{-1}),s) ) 
\end{equation}
and the leading asymptotic may cancel. The resulting asymptotic is then determined
by how many of the coefficients in the low temperature expansion of $Z^+_{S_N,s}(x_1,\cdots,x_d)$ and $Z^-_{S_N,s}(x_1,\cdots,x_d)$ are the same.

Note that even permutations form the subgroup $A_N$ of $S_N$ and $\vert A_N\vert=\frac{N!}{2}$ so that $\vert S_N\vert=2\vert A_N\vert$ and
$Z_{A_n}(x_a,\cdots,x_d,s)=2Z^{+}_{S_N}(x_1,\cdots,x_d,s)$ so that $Z_{A_N}(x_1,\cdots,s_d,s)$ will always satisfy an inversion relation.

We now specialise to the case $d=1$. For $ s N^{ s-1}$ odd, i,e. when $s,N$ are both odd, 
\begin{equation}
Z_{S_N}(x ,s)= (-1)^{N^s} x^{-N^s} Z_{S_N}(x^{-1} ,s) = (-1)^N Z_{S_N}(x^{-1} ,s)
\end{equation}
When $ sN^{ s-1}$ is even, i.e. one of $s , N $ are even then 
\bea 
Z_{S_N}(x,s)&& =(-1)^{ N^s}(x)^{-N^s} ( Z^{+}_{S_N} ( x^{-1} ,s)-Z^{-}_{S_N}(x^{-1},s) ) \cr 
&& = (-1)^{ N^s}(x)^{-N^s}  ( 2 Z^{+}_{S_N} ( x^{-1} ,s) - (  Z^{+}_{S_N} ( x^{-1} ,s) + Z^{-}_{S_N}(x^{-1},s) ) \cr 
&& = (-1)^{ N^s}(x)^{-N^s} ( Z_{ A_N } ( x^{-1} , s ) - Z_{ S_N} ( x^{-1} , s ) ) 
\eea 
When $x$ is taken to infinity, we can use the low-temperature expansion can be used for 
$ Z_{ A_N } ( x^{ -1} , s ) $ and $Z_{ S_N} ( x^{-1} , s )  $. Let $ c^{(1)}_{ S_N ; A_N ; s}  ( x^{-1} )^{ \Delta ( S_N ; A_N ; s ) } $ be the first non-zero difference when the RHS is expanded in powers of $x^{-1}$. Then the large $x$ behaviour is 
\bea\label{largexExps}  
Z_{S_N}(x,s) \sim (-1)^{ N^s}(x)^{-N^s - \Delta ( S_N ; A_N ; s ) }  c^{(1)}_{ S_N ; A_N ; s}
\eea 
We give  explicit examples of $S_N$ and $A_N$ partition functions for general $s$ and low $N$ in Appendix \ref{sec:examples}.

\subsection{ Behaviour at $ x \rightarrow 1 $  for large $N$ and general $s$ } 
\label{subsec:highT} 

Computations with Mathematica and the analysis of special families give strong evidence that the leading contributions in the high temperature limit $ x \rightarrow 1 $ come from $ p = [ 1^N ] $ followed by $ p = [ 2,1^{ N-2} ] $.

Start with the formula 
\bea 
\cZ_s ( N , p , x ) = 
\prod_{ i_1 , i_2 \cdots , i_s \in \{ 1, \cdots , N \} } 
{ 1 \over ( 1 - x^{ L ( a_{ i_1} , \cdots , a_{ i_s } )  } )^{  a_{ i_1} , \cdots , a_{ i_s } 
 p_{ i_1} \cdots p_{ i_s }  \over L ( a_{ i_1} , \cdots , a_{ i_s } )  } }
\eea

The degree function which determines how significantly a partition $p$ contributes as $ x \rightarrow 1$ is the power of $ ( 1- x ) $. Write $ p = [ a_1^{ p_1} , a_2^{ p_2} , \cdots , a_K^{ p_K } ] $ with $ p_1 , \cdots , p_K \ge 1 $ and $ \sum_{ i } a_i p_i = N$. 
\bea 
&& {\rm Degree } ( [ p , N , s ] )   = 
\sum_{ i_1 , \cdots , i_s =1}^{  K } { 1 \over L ( a_{ i_1} , \cdots , a_{ i_s}  ) } a_{ i_1 } \cdots a_{ i_s}   p_{ i_1} \cdots p_{ i_s} \cr 
&& = \sum_{ \substack {  S \subset \{ 1 , \cdots , K \} \\ S \ne \emptyset }  } 
     \sum_{ \substack {  H  \subset S \\ H \ne \emptyset } } { 1 \over L ( S ) }  (-1)^{ |S| - |H| } 
     (  \sum_{ i \in H  } a_{ i  } p_{ i  } )^s 
\eea
These equations follow from the expressions in Theorem 1 and Theorem 2 for $\cZ_s ( N , p , x )$.

First consider $ p = [ 1^N ] $, equivalently 
\bea 
&& a_1 = 1 \cr 
&& p_1 = N 
\eea
In this case, the only contribution to the product comes from the case 
$i_1 =1 , i_2 = 1 , \cdots , i_s = 1$, in which case $ p_{ i_1} = p_{ i_2} = \cdots = p_{ i_s } = N $. This gives 
\bea 
\cZ_m ( N , p = [1^N]  , x ) = { 1 \over ( 1 - x )^{ N^s } } 
\eea
Next consider $ p = [1^{ N-2} , 2 ] $, in which case 
\bea 
&& a_1  = 1 \cr 
&& a_2 =  2 \cr 
&& p_1 = N-2 \cr 
&& p_2 = 1
\eea
When all the $i_1 \cdots i_s $ take the value $1$, then we have 
\bea 
L ( 1,  1, \cdots , 1 ) = 1 
\eea
and 
\bea 
p_{ i_1} \cdots p_{ i_s } =  p_1^s = ( N-2)^s
\eea
When some of them ( a non-zero number)  take the value $2$ and remaining $1$, for example $k$ of them take value $1$ and remaining $m-k$ take value  $2$ with $ k \in \{ 0 , 1, \cdots , s-1 \}$. 
\bea 
L ( 1,  1, \cdots , 1   , 2 \cdots 2 ) = 2 
\eea
\bea 
{a_{ i_1 } \cdots a_{ i_s } \over L ( a_{i_1} , \cdots , a_{ i_s } ) } 
= {  1^{ k} 2^{ s-k} \over 2 } = 2^{ s-k-1} 
\eea
and 
\bea 
p_{ i_1} \cdots p_{ i_s} = ( N-2)^{ s} 
\eea
In this case the product evaluates to 
\bea 
{ 1 \over ( 1 - x^2 )^{ \sum_{ k =0}^{ s-1}  (N-2)^{ k} 1^{ s-k} {s \choose k } } }
\eea
The sum in the exponent is 
\bea 
{ 1 \over 2 } \sum_{ k =0}^{ s-1}  (N-2)^{ k} 2^{ s-k} {s \choose k }  = { 1 \over 2 } ( 
( ( N-2) +2 )^s - (N-2)^s )  = { N^s\over 2 }  -  { (N -2)^{ s} \over 2 } 
\eea 
\bea 
\cZ_s ( N , p = [1^{N-2} , 2]  , x ) = { 1 \over ( 1 - x )^{ (N-2)^{s   } }}
 { 1 \over ( 1 - x^2)^{{ N^s \over 2 }  -  { (N -2)^{ s} \over 2 } }}
\eea
Collecting the contributions from $p= [1^N] $ and $ p = [ 2,1^{N-2}]$
\bea 
\cZ ( N , x ) && = { 1 \over N! } \left (  { 1 \over ( 1-x)^{ N^ s } } 
+ { N ( N -1) \over 2 } { 1 \over ( 1 - x )^{ (N-2)^{ s   } }}
 { 1 \over ( 1 - x^2)^{{ N^s \over 2 }  -  { (N -2)^{ s } \over 2 } }}
 \right )  \cr 
 && =  { 1 \over N! }{ 1 \over ( 1-x)^{ N^ s } }   \left ( 1 + {N ( N -1) \over 2 } 
 {{{ ( 1- x )^{{ N^s \over 2 }  -  { (N -2)^{  s } \over 2 } }} \over ( 1 + x )^{{ N^s \over 2 }  -  { (N -2)^{  s } \over 2 } }} } \right ) 
\eea
Setting the second term to an $N$-independent constant $a$, we have 
\bea\label{bkdwnEq} 
{N ( N -1) \over 2 } {{{ ( 1- x )^{{ N^s \over 2 }  -  { (N -2)^{ s } \over 2 } }} \over ( 1 + x )^{{ N^s \over 2 }  -  { (N -2)^{ s } \over 2 } }} } = a 
\eea
we have at large $N$ 
\bea 
\log { N^2 \over 2 }  + { 1 \over 2 } ( N^s  - ( N-2)^s ) \log {  ( 1 - x ) \over ( 1 + x ) }  
= \log a 
\eea
This simplifies to 
\bea\label{leadingapprox}  
x \sim { \log N \over s N^{ s-1} }  + \hbox { terms sub-leading at large } ~ N 
\eea
This agrees with the matrix case 
\bea 
x \sim { \log N \over 2 N^{ } }  + \hbox { terms sub-leading at large } ~ N 
\eea
and generalises it to higher tensors. 

Note that $s=1$ is a special case, where there is no transition and there is no breakdown of the high temperature expansion associated with such a transition. 

It is  useful to give more explicit formulae for the breakdown point, expanded to subleading orders at large $N$. From \eqref{bkdwnEq} 
\bea 
\log \left (  { N^2 \over 2 } ( 1 - { 1 \over N } ) \right )  +  
\left ( {{ N^s \over 2 }  -  { (N -2)^{ s} \over 2 } }  \right ) \log { ( 1 - x ) \over ( 1 +x  ) } = \log a 
\eea
This simplifies to 
\bea 
\log { ( 1 - x ) \over ( 1 +x  ) } 
=  \left ( {{ N^s \over 2 }  -  { (N -2)^{ s} \over 2 } }  \right )^{-1}  \left (  \log a - \log \left (  { N^2 \over 2 } ( 1 - { 1 \over N } ) \right )  \right ) 
\eea
It is convenient  to define 
\bea 
&& F ( a , N ) = \left ( {{ N^s \over 2 }  -  { (N -2)^{ s} \over 2 } }  \right )^{-1}  \left (  \log a - \log \left (  { N^2 \over 2 } ( 1 - { 1 \over N } ) \right )  \right )  \cr 
&& = { 2 \over N^s }  \left ( 1 - ( 1 - {2 \over N })^{ s } \right )^{-1} 
  \left (  - \log ( N^2 /2 ) + \log a - \log (   1 - { 1 \over N } )      \right )  \cr 
    && = { 1 \over s N^{ s-1} } \left ( 1 +   \sum_{ l =2}^{ s }   { s! (-1)^{ l+1 }  \over l! ( s -l) ! } ( {  2 \over N } )^l  \right )^{-1} \left (  -  2 \log ( N  ) + \log (2a ) +   \sum_{ l=1}^{ \infty }  { N^{ - l }  \over l }  )      \right ) \cr 
    && = { - 2 \log N \over s N^{ s-1} } ( 1 + \cO ( 1/N ) ) 
        + { \log ( 2 a ) \over s N^{ s-1} }  ( 1 + \cO ( 1/N )  ) 
\eea 
The breakdown equation \eqref{bkdwnEq}  becomes 
\bea 
\log { ( 1 - x ) \over ( 1 +x  ) }  = F \, .  
\eea
This gives 
\bea 
x_{ \rm { bkdwn} } &&  = { ( 1 - e^F ) \over ( 1 + e^F ) }  = - \tanh ( F/2 )  \cr 
&& =  - \sum_{ n=1}^{ \infty } { 2^{ 2n} ( 2^{ 2n} -1 ) B_{ 2n} (F/2)^{ 2n-1}   \over ( 2n)! }
\eea
where we have used the expansion of the hyperbolic tangent in terms of Bernoulli numbers $B_{2n}$. The $n=1$ term in the above gives  the terms 
\bea 
 {  \log N \over s N^{ s-1} } ( 1 + \cO ( 1/N ) )  + { \log ( 2 a ) \over s N^{ s-1} }  ( 1 + \cO ( 1/N )  ) 
\eea
in agreement with the leading term \eqref{leadingapprox} 
The $n >1$ terms involve higher powers of $ {  \log N \over s N^{ s-1} } $ multiplied by higher powers of $ { \log ( 2 a ) \over s N^{ s-1} }$. 

\vskip.2cm 

\noindent
It is worth observing that the  critical Boltzmann factor $ x_c \sim { \log N  \over s N^{ s-1} } $ based on writing $ x = e^{ - \beta } $ (implicitly setting the hamonic oscillator mass and spring constant to $1$)  with $ \beta = { 1 \over T } $ implies 
\bea 
&& - { 1 \over T_c } \sim  \log (  { \log N \over s N^{ s-1} }  )  \sim - (s-1) \log N ~~\hbox{ at large }  N \cr 
&& \implies T_c \sim  { 1 \over ( s -1 ) \log N } 
\eea
which is a vanishing Hagedorn temperature. If we write the more detailed formula for  harmonic oscillators $ x = e^{ - \beta \omega } $ where $ \omega = \sqrt { k_{ \rm spring } \over m }  $ 
and treat $m,T$  as fixed and  $  k_{ \rm spring } $  as an experimentally tunable parameter, we have  a critical  $k_{ \rm spring ; c } $ given by 
\bea 
&& - { 1 \over T} \sqrt { k_{ \rm spring ; c } \over m }  \sim - (s-1) \log N \cr 
&& \implies  k_{ \rm spring ; c } \sim  { ( s-1)^2 m T^2  } ( \log N )^2  
\eea  
This cross-over spring constant for finite $m, T$ and large finite  $N$  may, speculatively be amenable to experimental realisation. Also note that scaling $k_{{\rm spring},c}$ as ${\bar k}_{{\rm spring}, c}(\log  N)^2 $ yields an $N$ independent critical value and equivalently a finite critical temperature $T_c=\frac{1}{s-1}\sqrt\frac{{\bar k}}{m}$.

\section{ Negative heat capacities from near-factorial degeneracies, scaling limits and thermodynamics  }\label{sec:thermodynamics} 

There is a rich thermodynamics, involving a zero-temperature Hagedorn transition, negative specific heat capacities and inequivalence between micro-canonical and canonical ensembles, encoded in the thermodyamic partition function for the $s=2$ case, which was studied in detail in \cite{PIMQM-Thermo}.  Related recent observations in the context of the statistical thermodynamics of unlabeled networks are in \cite{Evnin} while earlier discussions of negative specific heat capacities with string theoretic motivations related to the present work are in \cite{Hanada,Ber1,Barbon,Aki}. These were related to the fact that the coefficients in the small $x$ expansion of $ \cZ_{s=2}  ( N , x ) $ grow near-factorially as a function of the  degree 
 and then are tamed when this degree becomes large compared to $ N \log N$.  In this section we initiate the extension of the detailed study for the case $ s > 2 $. To set this up, we describe an infinite family of near-factorial degeneracies which lead to negative specific heat capacities in the micro-canonical ensemble. This is followed by illustrative plots which indicate that the thermodynamics in the higher $s$ case can be expected to be very similar to the $s=2$ case previously studied.

\subsection{  A family of near-factorial degeneracies with negative heat capacity  } 
\label{subsec:nearfac}

As observed in \cite{PIMQM-Thermo} large $k$  degeneracies of the form  $ \Omega ( k ) = k! $ and corresponding 
$ S ( k ) = k \log k $ have 
\bea 
&& S' ( k ) = 1 + \log k \cr
&& S'' ( k ) = 1/k 
\eea
The positivity of $ S'' ( k ) $ gives negative heat capacity in the micro-canonical ensemble. 
Consider more generally antropy of the form 
\bea 
S ( k ) = k^ a ( \log k )^b f ( k )  
\eea
with $ a \ge 1 , b \ge 0 $ and 
\bea 
f(k) \rightarrow 1 , f'(k) \rightarrow 0 , f'' (k ) \rightarrow 0 
\eea 

The second derivative of the entropy is 
\bea 
&& S'' ( k) = 
k^{a-2} (\log k )^{ b-2}   \bigg( ~ 2 f'(k) k \log k (a \log k +b)+ (k \log k )^2 f''(k)~  \cr 
&& 
 +f(k) \big( ~ (2 a-1) b \log k +(a-1) a ( \log k )^2 +(b-1) b~ \big)\bigg) \cr 
 && 
\eea 
Assuming that $ a > 1  $, no assumptions on $b$,  
\bea 
&& f(k) \rightarrow 1 \cr 
&& f'(k ) k \rightarrow 0 \cr 
&& f'' ( k ) k^2  \rightarrow 0 
\eea 
in the large $k$ limit, the dominant term in that limit is 
\bea 
 S'' ( k) \sim a (a-1) k^{a-2} (\log k )^{ b}  f ( k ) 
\eea
with positive second derivative, and hence negative heat  capacity. 

Assuming $ a =1, b > 0  $, and 
\bea 
&& f(k ) \rightarrow 1 \cr 
&& f'(k ) k \log k  \rightarrow 0 \cr 
&& f'' ( k ) k^2   \log k \rightarrow 0 
\eea
in the large $k$ limit, the dominant term in that limit is 
\bea 
 S'' ( k) \sim b  k^{-1} ( \log k )^{ -1} ]  f ( k )  
\eea 
with positive second derivative, and hence negative heat  capacity. 

Assuming $ 0 < a < 1$, no assumption on $b$, but assuming 
\bea 
&& f ( k ) \rightarrow 	1 \cr 
&& k f'( k ) \rightarrow c  \cr
&& c > { ( 1- a ) \over 2 } \cr  
&& k^2 f'' ( k ) \rightarrow 0 
\eea
the leading term in the second derivative is 
\bea 
S'' ( k ) = a ( 2 c - ( 1 - a ) ) k^{ a -2 } \log k 
\eea
which is positive, implying negative heat capacity. 

We have therefore characterized an infinite family of large $k$ behaviours of the micro-canonical entropy, which are in an appropriately defined sense, near $ S ( k ) \sim k \log k $, and which lead to negative heat capacity. 

\subsection{ Thermodynamics  } 

Given the genericity of negative heat capacities for super-factorial and near-factorial degeneracies, it is reasonable to expect that the thermodynamic features observed in \cite{PIMQM-Thermo} are generic. These features are negative heat capacity determined by the rapid growth of states in the stable range of energies, in-equivalence of ensembles since the heat capacity in the canonical ensemble is always positive,  a sharp transition in the canonical ensemble getting sharper as $ N \rightarrow \infty$, 

The example plots of energy vs temperature in the canonical and micro-canonical ensemble support this expectation. We will return to a more systematic discussion of the characteristics of the transition for the $ s > 2 $ case in the future. 

\begin{figure}
\includegraphics[scale=0.5]{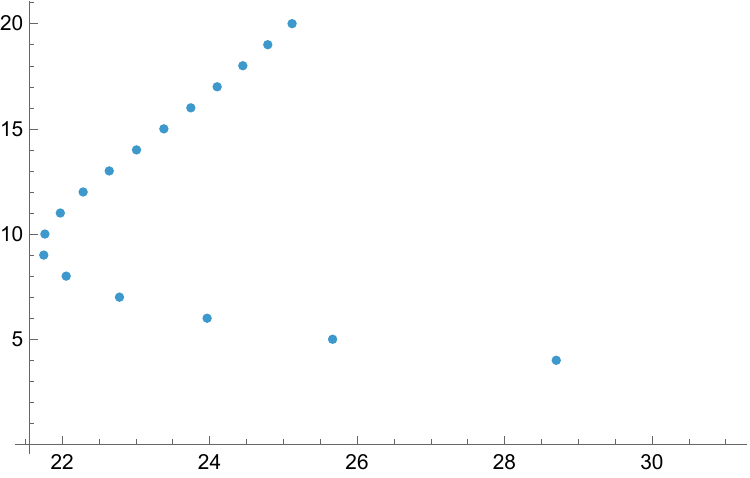}
\caption{Micro-canonical energy versus temperature $ s=3$, $ N = 10$ } 
\label{EvsTMicro-seq3-Neq10} 
\end{figure} 

\begin{figure} 
\includegraphics[scale=0.5]{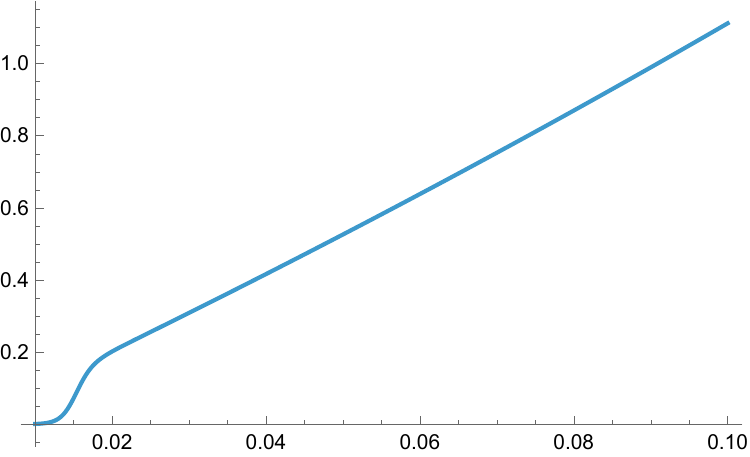}
\caption{Canonical energy versus temperature $ s=3$, $ N = 10$ } 
\label{EvsTCanonical-seq3-Neq10} 
\end{figure} 

\section{ Discussion : Future directions }

The present work opens up a number of interesting questions for future investigation. 
In section \ref{limsdlts} we conjectured, based on ample evidence from analytical formulae and computations in Mathematica,  that the leading two terms in the high temperature expansion of 
$ \cZ_s ( N , p , x ) $ for general $s$ come from $ p = [1^N ] $ followed by $ [ 2,1^{ N-2} ] $. 
For the case $s=2$, this  was rigorously proved in Appendix B  of  \cite{PIMQM-Thermo}. The general $s$ conjecture was used here to derive the leading large $N$ form of the critical Boltzmann factor $ x_c \sim { \log N \over s N^{ s- 1 } } $, where the high temperature expansion breaks down. It would be very interesting to find the proof of the general $s$ conjecture for the leading two terms. 

\vskip.2cm 

The tensorial harmonic oscillator system we considered in this paper has the simplest permutation invariant quadratic potential. The most general refined  tensor harmonic oscillator allows
general non-zero  parameters for the distinct linearly independent quadratic invariants built from the tensor variables. The computation of correlators for refined tensor models having general quadratic parameters was described in \cite{PITM}. The Molien-Weyl formula was found to be an effective tool in computing the canonical ensemble partition function for the refined matrix harmonic oscillators \cite{GPIMQM-PF}. Extending the calculation of tensor partition functions obtained here to the refined case is an interesting direction. 

\vskip.2cm 

A third direction is to use the formulae developed here to characterise more precisely the stable counting regions for $S_N$ and $A_N$ partition functions, to develop the graph counting interpretations of the small $x$ expansions (along the lines of the results in \cite{PIGMM}\cite{PIG2MM}\cite{PITM}), and find precise exponents for  the large $x$ behaviours using equation \eqref{largexExps} in section \ref{limsdlts}. 

\vskip.2cm 

This paper has uncovered some unexpected links between the statistical thermodynamics of gauged permutation invariant tensor models and  basic number-theoretic structures at play with combinatorial structures in an intriguing way,  notably least common multiples  of subsets of parts of partitions, with important simplifications resulting from the use of the inclusion-exclusion principle of combinatorics. The theme of deep connections between interesting statistical thermodynamics and number-theoretic structures uncovered here is likely to be part of a wider story likely including other recent results such as \cite{Fink,SheldonFink}, where LCMs of parts of a partition also play an important role in the context Kauffmann models of statistical physics \cite{Kauffman1,Kauffman2}.

\begin{center} 
{\bf Acknowledgements } 
\end{center} 

It is a pleasure to acknowledge useful conversations with  Joseph Ben Geloun, David Berenstein,  Robert de Mello Koch,   Thomas Fink, Yang-Hui He, Vishnu Jejjala, Yang Lei, Forrest Sheldon,  Bo Sundborg.  SR is supported by the Science and Technology Facilities Council (STFC) Consolidated Grant ST/T000686/1 
``Amplitudes, strings and duality''. SR grateful for a Visiting Professorship at Dublin Institute for Advanced Studies,  held during 2024, where this project was initiated.  SR also gratefully acknowledges a visit to the Perimeter Institute in November 2024: this research was supported in part by Perimeter Institute for Theoretical Physics. Research at Perimeter Institute is supported by the Government of Canada through the Department of Innovation, Science, and Economic Development, and by the Province of Ontario through the Ministry of Colleges and Universities.

\begin{appendix} 

\section{ Mathematica code for implementing the formula in Theorem 2.   } 

The following mathematica code calculates the canonical partition functions for symmetric groups $S_N$ and and alternating groups  $A_N$. 
\begin{verbatim} 
(* symmetry factor *)
Sym [ p_  , N_ ] := 
 Product [  i^(Count [ p , i ]) Factorial [  Count [ p , i ] ] , { i , 1 , N }]
(* Subset sum formulae for partition functions *)
ZZsubsets [P_List, s_, x_] := 
 Module[{p, K, aValues, pValues, indexSubsets, factors, A, B, result},
   p = Tally[P];(*Convert partition to {{part,multiplicity},...}*)
  K = Length[p];
  aValues = p[[All, 1]];
  pValues = p[[All, 2]];
  indexSubsets = Subsets[Range[K], {1, K}];(*all non-empty subsets*)
  factors = 
   Table[Module[{S = subset, subsetsOfS}, A = LCM @@ aValues[[S]];
     subsetsOfS = Subsets[S, {1, Length[S]}];(*exclude empty subset*)
     B = Total[(-1)^(Length[S] - Length[#])*1/
           A*(Total[(aValues[[#]]*pValues[[#]])])^s & /@ subsetsOfS];
     (1/(1 - x^A))^B], {subset, indexSubsets}];
  result = Times @@ factors;
  result]
 (* Canonical partition function for $S_N$ *) 
ZZsubsetAns [ N_ , s_ , x_ ] := 
 Sum  [1/  Sym [ p  , N ] *  ZZsubsets [ p , s, x] , { p , 
   IntegerPartitions [ N ]}] ; 
 ZZsubsetAltntng [ N_ , s_ , x_ ] :=  ( 
  ZZsubsetAns [ N, s , x ]  +  ZZsubsetAns [ N, s , x^(-1) ] ) 
\end{verbatim} 
The computation of the $S_N$ partition functions for different values of $N$ and general $s$ are performed by implementing using the function 
\begin{verbatim}
 ZZsubsetAns [ N_ , s_ , x_ ],
\end{verbatim} while the computation of $A_N$ partition functions is done by the function 
\begin{verbatim}
 ZZsubsetAltntng [ N_ , s_ , x_ ]. 
\end{verbatim} 

\section{ Examples of $S_N$ and $A_N$ partition functions   }\label{sec:examples} 

Example partition functions calculated using \eqref{MainProps2} are 
\begin{equation}
Z_{S_N,1}(x)=\frac{1}{\prod_{n=1}^N(1-x^n)}\quad\hbox{ and}\quad
\frac{Z_{A_N,1}(x)}{Z_{S_N,1}(x)}=1+x^{\frac{N(N-1)}{2}}
\end{equation}
There is one new invariant, the Vandermonde of the vector components
which is invariant under $A_N$ but is not invariant under $S_N$.

\begin{equation}
  Z_{S_2}(x,s)=\frac{1}{2}\frac{1}{(1-x)^{2^s}}+\frac{1}{2}\frac{1}{(1-x^2)^{\frac{2^s}{2}}}\quad\hbox{ and}\quad
 Z_{A_2}(x,s)=\frac{1}{(1-x)^{2^s}}
\end{equation}
\begin{eqnarray}
  Z_{S_3}(x,s)&=&\frac{1}{3!}\frac{1}{(1-x)^{3^s}}+\frac{1}{2}\frac{1}{(1-x)(1-x^2)^{\frac{3^{s}-1}{2}}}+\frac{1}{3}\frac{1}{(1-x^3)^{\frac{3^{s}}{3}}}\nonumber\\
  &&\\
  Z_{A_3}(x,s)&=&\frac{1}{3}\frac{1}{(1-x)^{3^s}}+\frac{2}{3}\frac{1}{(1-x^3)^{\frac{3^{s}}{3}}}
  \nonumber
\end{eqnarray}
\begin{eqnarray}
  Z_{S_4}(x,s)&=&\frac{1}{4!}\frac{1}{(1-x)^{4^s}}+\frac{1}{8}\frac{1}{(1-x^2)^{\frac{4^{s}}{2}}}+\frac{1}{4}\frac{1}{(1-x)^{2^s}(1-x^2)^{\frac{4^{s}-2^s}{2}}}\hfill\nonumber\\
  &&\qquad+\frac{1}{3}\frac{1}{(1-x)(1-x^3)^{\frac{4^{s}-1}{3}}}+\frac{1}{4}\frac{1}{(1-x^4)^{\frac{4^{s}}{4}}}\nonumber\\
  &&\\
  Z_{A_4}(x,s)&=&\frac{2}{4!}\frac{1}{(1-x)^{4^s}}+\frac{1}{4}\frac{1}{(1-x^2)^{ \frac{4^s -1 }{2}}}+\frac{2}{3}\frac{1}{(1-x)(1-x^3)^{\frac{4^{s}-1}{3}}}
  \nonumber
\end{eqnarray}

\begin{eqnarray}
  Z_{S_5}(x,s)&=&\frac{1}{5!}\frac{1}{(1-x)^{5^s}}+\frac{1}{8}\frac{1}{(1-x)(1-x^2)^{\frac{5^s-1}{2}}}+\frac{1}{12}\frac{1}{(1-x)^{3^s}(1-x^2)^{\frac{5^s-3^s}{2}}}\hfill\nonumber\\
  &&\qquad+\frac{1}{6}\frac{1}{(1-x)^{2^s}(1-x^3)^{\frac{5^{s}-2^s}{3}}}
  +\frac{1}{6}\frac{1}{(1-x^2)^{\frac{2^s}{2}}(1-x^3)^{\frac{3^s}{3}}(1-x^6)^{\frac{5^s-3^s-2^s}{6}}}\hfill\nonumber\\
  &&\qquad\qquad+\frac{1}{4}\frac{1}{(1-x)(1-x^4)^{\frac{5^{s}-1}{4}}}+\frac{1}{5}\frac{1}{(1-x^5)^{\frac{5^{s}}{5}}}\nonumber\\
 &&\\
  Z_{A_5}(x,s)&=&\frac{2}{5!}\frac{1}{(1-x)^{5^s}}+\frac{1}{4}\frac{1}{(1-x)(1-x^2)^{\frac{5^s-1}{2}}}+\frac{1}{3}\frac{1}{(1-x)^{2^s}(1-x^3)^{\frac{5^{s}-2^s}{3}}}+\frac{2}{5}\frac{1}{(1-x^5)^{\frac{5^{s}}{5}}}
  \nonumber
\end{eqnarray}

\begin{eqnarray}
  Z_{S_6}(x,s)&=&\frac{1}{6!}\frac{1}{(1-x)^{6^s}}+
  \frac{1}{48}\frac{1}{(1 - x^2)^{\frac{6^s}{2}}}
    + \frac{1}{16}\frac{1}{(1 - x)^{2^s}(1 - x^2)^{\frac{6^s - 2^s}{2}}}\nonumber\\
    &&\qquad + \frac{1}{48}\frac{1}{(1 - x)^{4^s} (1 - x^2)^{\frac{6^s - 4^s}{2}}}
    + \frac{1}{18}\frac{1}{(1 - x^3)^{\frac{6^s}{3}}}+ \frac{1}{18}\frac{1}{(1 - x)^{3^s} (1 - x^3)^{\frac{6^s - 3^s}{3}}}\nonumber\\
 &&\qquad\quad 
    + \frac{1}{6}\frac{1}{(1 - x) (1 - x^2)^{\frac{3^s - 1}{2}} (1 - x^3)^{\frac{4^s - 1}{3}} (1 - x^6)^{\frac{6^s - 4^s - 3^s + 1}{6}}}
  + \frac{1}{8}\frac{1}{(1 - x)^{2^s} (1 - x^4)^{\frac{6^s - 2^s}{4}}}\nonumber\\
 &&\qquad\quad + \frac{1}{8}\frac{1}{(1 - x^2)^{2^{s - 1}} (1 - x^4)^{\frac{6^s - 2^s}{4}}}
  + \frac{1}{5}\frac{1}{(1 - x) (1 - x^5)^{\frac{6^s - 1}{5}}}
  + \frac{1}{6}\frac{1}{(1 - x^6)^{6^{s - 1}}}
\end{eqnarray}

\begin{eqnarray}
  Z_{A_6}(x,s)&=&\frac{2}{6!}\frac{1}{(1-x)^{6^s}}
    + \frac{1}{8}\frac{1}{(1 - x)^{2^s}(1 - x^2)^{\frac{6^s - 2^s}{2}}}\nonumber\\
    &&\qquad + \frac{1}{9}\frac{1}{(1 - x^3)^{\frac{6^s}{3}}}+ \frac{1}{9}\frac{1}{(1 - x)^{3^s} (1 - x^3)^{\frac{6^s - 3^s}{3}}}\nonumber\\
 &&\qquad\quad + \frac{1}{4}\frac{1}{(1 - x^2)^{2^{s - 1}} (1 - x^4)^{\frac{6^s - 2^s}{4}}}
  + \frac{2}{5}\frac{1}{(1 - x) (1 - x^5)^{\frac{6^s - 1}{5}}}
\end{eqnarray}

\end{appendix}

\end{document}

\bibitem{RobertsTesman} 
Roberts, Fred S.; Tesman, Barry (2009), Applied Combinatorics (2nd ed.), CRC Press, ISBN 9781420099829

\bibitem{WikiLCM} Wikipedia article on least common multiples, 
\url{https://en.wikipedia.org/wiki/Least\_common\_multiple}
